\documentclass[twocolumn,11pt,twoside]{article}
\usepackage{epsfig}

\def\modell{{\sl poly-gonato} model}
\def\lna{\langle\ln A\rangle}

\def\knie{{\sl knee}}

\begin{document} 
\twocolumn[{\centering
 \vspace*{3cm}
{\LARGE\bf Models of the Knee in the Energy Spectrum of Cosmic Rays\\[5mm]}

{\Large J\"org R. H\"orandel\\[3mm]}
{\small\sl University of Karlsruhe, Institut f\"ur Experimentelle Kernphysik,
            P.O. Box 3640, 76\,021 Karlsruhe, Germany\\
            http://www-ik.fzk.de/$\sim$joerg\\[5mm]}

Preprint arXiv:astro--ph/0402356\\
Submitted to Astroparticle Physics 21. November 2003; 
        accepted  26. January 2004\\[5mm]
}	

\hspace*{0.1\textwidth}\begin{minipage}{0.8\textwidth}
\hrulefill\\
{\bf Abstract\\}
{\small
The origin of the \knie\ in the energy spectrum of cosmic rays is an
outstanding problem in astroparticle physics.  Numerous mechanisms have been
proposed to explain the structure in the all-particle spectrum. In the article
basic ideas of several models are summarized, including diffusive acceleration
of cosmic rays in shock fronts, acceleration via cannonballs, leakage from the
Galaxy, interactions with background particles in the interstellar medium, as
well as new high-energy interactions in the atmosphere.  The calculated energy
spectra and mean logarithmic masses are compiled and compared to results from
direct and indirect measurements.
}

\hrule
\end{minipage}

\vspace*{5mm}
]

\section{Introduction}

Cosmic-ray particles --- fully ionized atomic nuclei --- are the only matter
which reaches the earth from outside the solar system and are accessible for
direct investigations. The energy of these particles can exceed even
$10^{20}$~eV, but their acceleration mechanisms and the sites of origin are
still under discussion.  The energy spectrum is of non-thermal origin and
follows a power law $dN/dE\propto E^{\gamma}$ over many orders of magnitude.
The spectrum steepens at energies around 4~PeV from a spectral index
$\gamma\approx-2.7$ to $\gamma\approx-3.1$. This feature is commonly called the
\knie\ and its explanation is generally believed to be a corner stone in
understanding the origin of cosmic rays, providing answers to one of the key
questions of astroparticle physics.

Many approaches have been discussed in the literature to describe the origin,
acceleration, and propagation of cosmic rays and several mechanisms have been
proposed to explain the \knie. In this article the basic ideas of theoretical
models are briefly mentioned and the resulting energy spectra and mass
compositions are described in section~\ref{theosect}.  

The steeply falling cosmic-ray flux as function of energy requires increasing
apertures and exposure times. Presently, above 1~PeV cosmic rays are
experimentally accessible in ground-based installations only. These detectors do
not measure the primary particles, instead secondary particles are measured,
produced in high-energy interactions in the atmosphere, forming an extensive
air shower.  This makes the interpretation of the indirect measurements very
difficult and the results obtained depend on the understanding of high-energy
interactions in the atmosphere.  The actual experimental status is briefly
discussed in section~\ref{pgsect}.

\section{Experimental results and the \modell} \label{pgsect}

Recently, the results of direct measurements of energy spectra of individual
elements and indirect measurements of the all-particle spectrum of cosmic rays
have been investigated by the author \cite{polygonato}. It has been shown, that
the observed energy spectrum and mass composition can be described consistently
in the energy range from 10~GeV up to at least 100~PeV, adopting a
phenomenological model, the \modell.  

The energy spectra of individual nuclei are assumed to follow power laws with a
cut-off at high energies.  For the cut-off different approaches have been
tried, the cut-off energy has been taken proportional to either the nuclear
charge or the nuclear mass of the individual elements. In addition, also a
constant cut-off energy has been assumed.  Two possibilities for the spectral
shape above the cut-off energy have been investigated, a common spectral slope
$\gamma_c$ for all elements and a constant difference $\Delta\gamma$ of the
spectral index before and above the cut-off energy.

Individual spectra of elements, as directly measured at the top of the
atmosphere for energies below 1~PeV, are extrapolated to high energies and
fitted to results from air-shower measurements.  For these indirect
measurements the individual energy spectra had been normalized to match the
all-particle spectrum of direct measurements at 1~PeV. It turned out, that only
small energy shifts were necessary, all within the quoted errors --- but as a
result all air-shower measurements yield consistent all-particle energy
spectra.  

It turned out that a cut-off proportional to the nuclear charge describes the
data best. The modeling of the cut-off behavior seems not to be crucial for
the description of the measurements, the two assumptions of a common $\gamma_c$
and a common $\Delta\gamma$ yield almost the same results. The energy spectra
obtained are hardly distinguishable from each other. 
In addition, a significant contribution of ultra-heavy elements in the energy
region around $10^8$~GeV has been found.
With the \modell\ two prominent features in the cosmic-ray spectrum can be
explained. The \knie\ at 4.5~PeV follows from subsequent cut-offs for
individual elements, starting with protons, and the {\sl second} \knie\ at
400~PeV is due to the end of the galactic component.

\begin{figure*}\centering
 \epsfig{file=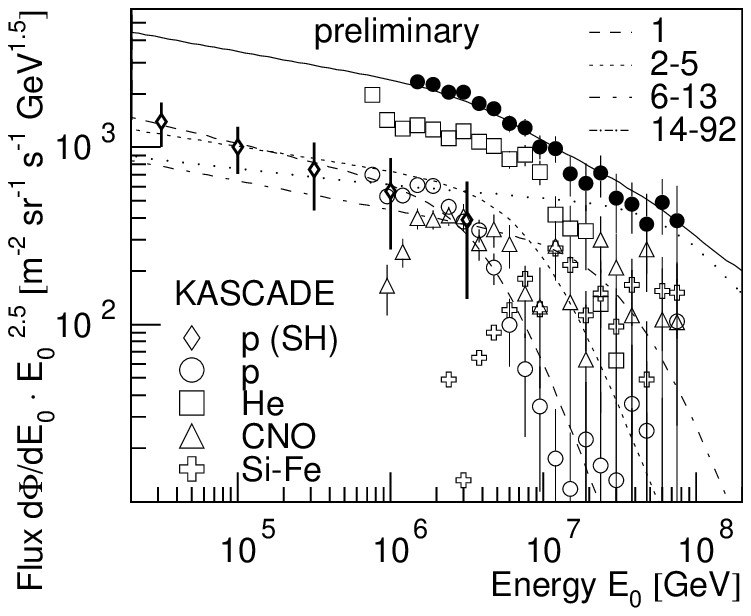,width=\columnwidth}
 \epsfig{file=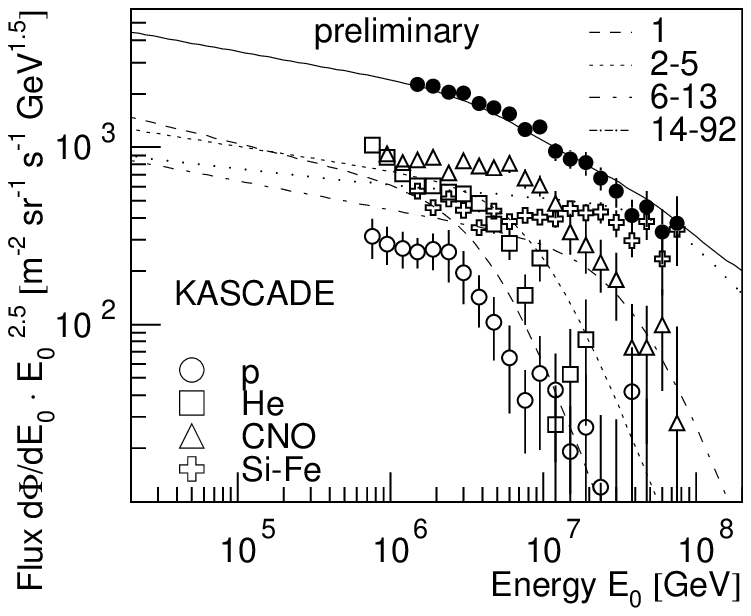,width=\columnwidth}
 \caption{Preliminary energy spectra for groups of elements as obtained by the 
	  KASCADE experiment investigating the electromagnetic and muonic
	  components \cite{taup}, as well as single hadrons \cite{mueller},
	  using CORSIKA simulations with the hadronic interaction models QGSJET
	  (left-hand side) and SIBYLL (right-hand side) to interpret the data.
	  The energy scale of the data has been readjusted, see text.  The
	  all-particle flux is represented by the filled symbols.  The solid
	  line gives the average all-particle flux from several experiments
	  \cite{polygonato}. The dashed and dotted lines indicate spectra
	  according to the \modell\ for elemental groups with nuclear charge
	  numbers as listed.}
 \label{kascade}	  
\end{figure*}

In the past air shower experiments mostly derived all-particle energy spectra
from their measurements.  However, recently also results on the energy spectra
for groups of elements in the PeV region have been derived from indirect
measurements. 

The KASCADE group performed systematic studies to evaluate the influence of
different hadronic interaction models used in the simulations to interpret the
data on the resulting spectra for elemental groups
\cite{ulrich,ulrichisvh,rothisvh,roth,taup}.  Two sets of preliminary spectra,
derived from the observation of the electromagnetic and muonic air shower
components, using CORSIKA \cite{corsika} with the hadronic interaction models
QGSJET and SIBYLL are compiled in Figure~\ref{kascade}.  In order to compare
the results with other measurements and theoretical predictions, the flux of
the all-particle spectrum has been normalized to the average measured flux, as
obtained by many experiments \cite{polygonato}, represented by the solid line
in the figure.  For the normalization, the energy scale has been slightly
adjusted by $\delta_E=-15\%$ for both interaction models used.  For comparison,
also spectra according to the \modell\ are given.

As can be seen in the figure, the flux for elemental groups depends on the
model used.  The KASCADE group emphasizes that at present there are systematic
differences between measured and simulated observables, which result in the
ambiguities of the spectra for elemental groups. These conclusions apply
similarly also to other experiments.  A correct deconvolution of energy spectra
for elemental groups requires a precise knowledge of the hadronic interactions
in the atmosphere.  It seems that the interaction models presently used do not
describe the measurements with the high precision required.  

Nevertheless, analyzing the flux for individual elemental groups, both sets of
spectra indicate a rigidity dependent cut-off.  While the absolute flux values
obtained are different compared to the \modell, the cut-off behavior for the
individual groups is very similar to the spectra according to the
phenomenological approach.  Also presented is the flux of protons obtained from
an analysis of unaccompanied hadrons \cite{mueller}, which is found to be
compatible with the extrapolations of the direct measurements below the \knie.

\begin{figure}\centering
 \epsfig{file=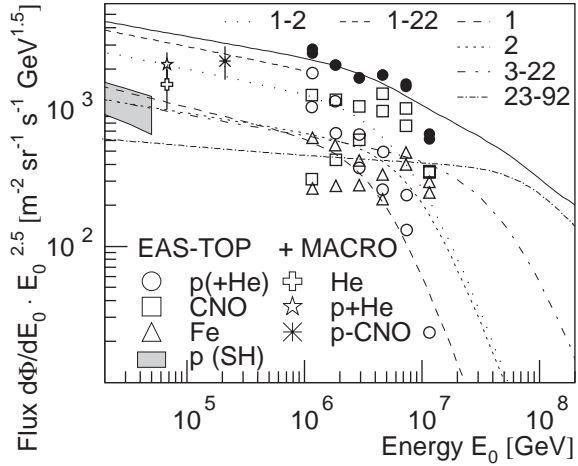,width=0.95\columnwidth}
 \caption{Energy spectra for groups of elements as obtained by the
	  EAS-TOP experiment from the analysis of single hadrons
	  \cite{castellina}, the electromagnetic and muonic components
	  \cite{navarra,valchierotti}, and a combined analysis of \v{C}erenkov
	  light and high-energy muons registered by the MACRO detector
	  \cite{bertaina}.  The energy scale of the data has been readjusted,
	  see text.  The all-particle flux is represented by the filled
	  symbols, see also caption of Figure~\ref{kascade}.}
 \label{eastop}	  
\end{figure}

The results of the EAS-TOP experiment are summarized in Figure~\ref{eastop}.
The proton spectrum as derived from unaccompanied hadrons \cite{castellina}
agrees well with the extrapolation of the direct measurements.  The combined
analysis of EAS-TOP \v{C}erenkov and MACRO muon data \cite{bertaina} have been
normalized ($\delta_E=-15\%$) to the proton+helium flux of direct measurements,
otherwise the flux for p+He+CNO would exceed the average measured all-particle
spectrum.  The results obtained are compatible with the extrapolation of the
direct measurements. The spectra derived from the number of electrons and muons
(scaled by $\delta_E=-5\%$) are given for two cases
\cite{navarra,valchierotti}. In the first, the light component consists of
protons only and in the second scenario it is a mixture of protons and helium
nuclei. The range obtained is indicated by the two sets of points for each
mass group.

\begin{figure}\centering
 \epsfig{file=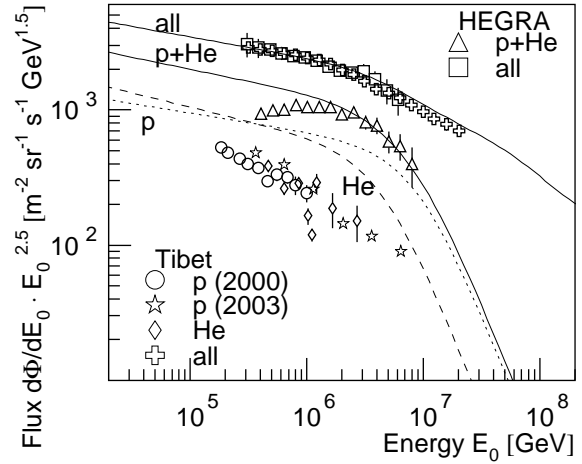,width=\columnwidth}
 \caption{Energy spectra for groups of elements as obtained by the HEGRA 
	  \cite{hegra} and Tibet \cite{tibet,tibet00,tibet03} groups.  The
	  energy scale of the data has been readjusted, see text.  The solid
	  line gives the average all-particle flux from several experiments
	  \cite{polygonato}.  The dashed and dotted lines indicate spectra
	  according to the \modell\ for protons and helium nuclei.}
 \label{hegra}	  
\end{figure}

The HEGRA collaboration derived a sum-spectrum for the proton and helium
component as well as an all-particle spectrum \cite{hegra}, as shown in
Figure~\ref{hegra}.  The all-particle spectrum has been normalized to the
average measured flux ($\delta_E=-10\%$). Below 1~PeV the flux for protons and
helium is lower than the extrapolations of the direct measurements, while at
higher energies the data compare well to the \modell.

The figure also reviews the spectra obtained by the Tibet group
\cite{tibet,tibet00,tibet03}, the all-particle spectrum has been normalized to
the average measured spectrum ($\delta_E=-10\%$). The spectra for protons and
helium nuclei are below the extrapolations of the direct measurements using
single power laws.  However, one has to keep in mind that these results are
obtained with a very limited number of events observed. As example, the results
of 2003 \cite{tibet03} are based on less than 200 registered events. On the
other hand, the experiments mentioned previously use data with high statistics,
e.g. the KASCADE group uses $O(10^6)$ events for their analysis.

It is worth mentioning that all energy shifts applied have a negative sign, i.e.
the energy is overestimated by the hadronic interaction models used to
interpret the measured data. A possible explanation for this behavior could be
a too large increase of the inelastic cross sections as function of energy, as
pointed out in \cite{wq}.

It should be emphasized that the \modell\ is based on individual element
spectra from direct measurements and the all-particle spectrum from indirect
measurements only. The energy spectra for elemental groups derived from air
shower measurements as just discussed are not taken into account. Hence, the
\modell\ can be treated as independent result and it is interesting to realize
that the predicted spectra are compatible with the deconvoluted spectra from
indirect measurements.

However, the present experimental status concerning energy spectra for
elemental groups in the PeV region is not yet conclusive.  Additional
information from a robust observable like the mean logarithmic mass would be
very helpful to evaluate the predictions of various theoretical models of the
\knie. The mean logarithmic mass, defined as $ \lna=\sum r_i \ln A_i$, with the
relative fraction $r_i$ of the element with mass number $A_i$ is an often used
quantity to characterize the cosmic-ray mass composition.  

In the earlier article \cite{polygonato} experimental $\lna$ values have been
investigated.  A systematic difference has been found in the mean logarithmic
mass calculated from experiments measuring particle distributions at ground
level and detectors observing the average depth of the shower maximum. A recent
investigation showed, that the discrepancy can be reduced, if the logarithmic
increase of the inelastic cross-sections in the model QGSJET is lowered and the
elasticity of the interactions is increased \cite{wq}. 

\begin{table}
\caption{Average mean logarithmic mass $\lna$ as obtained from many experiments
	 as function of primary energy $E_0$ [GeV].  $\sigma_{\lna}$ denotes
	 the variation, not the error of the mean value.} 
\renewcommand{\arraystretch}{1.0}
\begin{tabular}{lll} \hline
$\log_{10}E_0 [GeV]$ & $\lna$ & $\sigma_{\lna}$ \\ \hline
 5.12&1.54&0.37\\
 5.38&1.74&0.47\\
 5.62&1.81&0.44\\
 5.88&1.93&0.42\\
 6.12&1.86&0.49\\
 6.38&1.84&0.60\\
 6.62&1.94&0.68\\
 6.88&2.22&0.73\\
 7.12&2.54&0.62\\
 7.38&2.93&0.45\\
 7.62&3.17&0.44\\
 7.88&2.88&0.50\\
 8.12&2.84&0.43\\
 8.38&2.87&0.34\\
 8.62&2.85&0.37\\
 8.88&2.69&0.57\\
\hline
\end{tabular}
\label{massetab}
\end{table}

\begin{figure*}\centering
 \epsfig{file=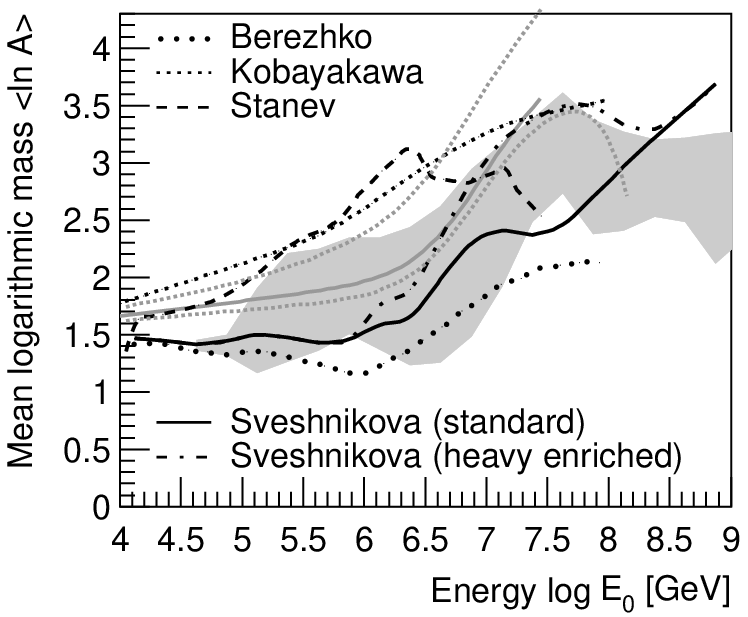,width=\columnwidth}
 \epsfig{file=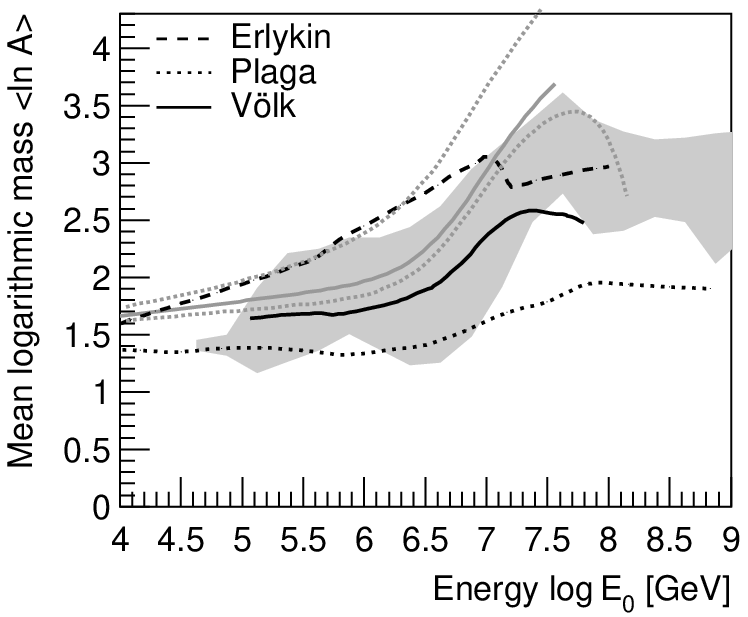,width=\columnwidth}
 \epsfig{file=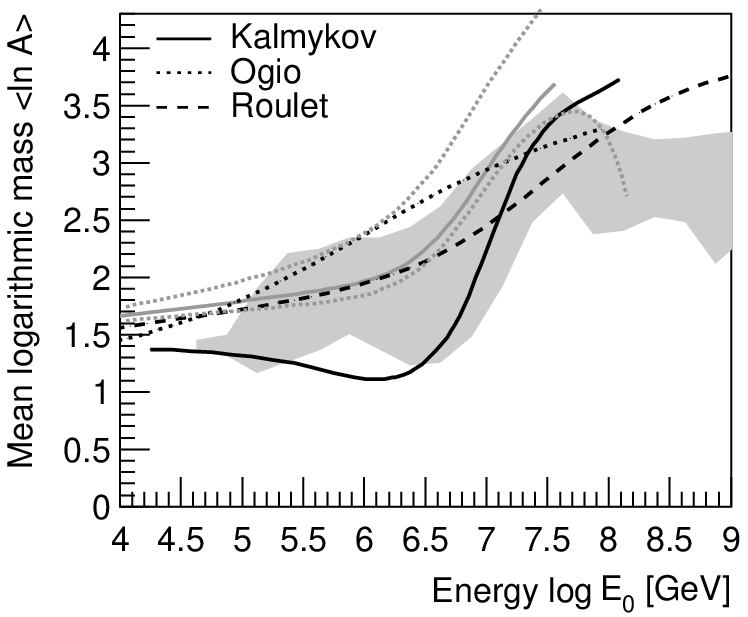,width=\columnwidth}
 \epsfig{file=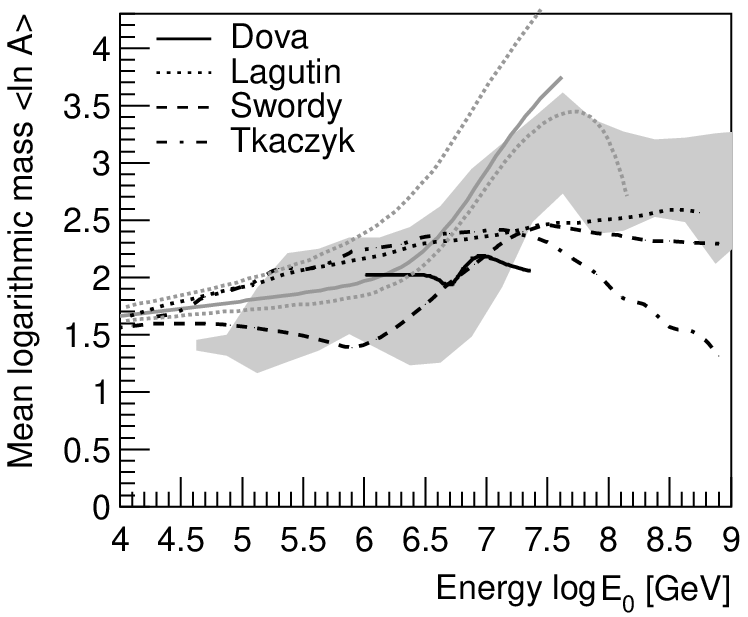,width=\columnwidth}
 \caption{Mean logarithmic mass derived from many experiments (shaded area) 
	  compared to $\lna$ as obtained with different models.
	  Upper left panel: Acceleration in supernova remnants as described by
	  Berezhko and Ksenofontov \cite{berezhko},
	  Kobayakawa et al. \cite{kobayakawa},
	  Stanev et al. \cite{stanev}, and
	  Sveshnikova \cite{sveshnikova}.
	  Upper right panel: Source and acceleration related models by
	  Erlykin and Wolfendale \cite{wolfendale},
	  Plaga \cite{plaga}, as well as
	  V\"olk and Zirakashvili \cite{voelk}.
	  Lower left panel: Diffusion models by
	  Kalmykov et al. \cite{kalmykov},
	  Ogio and Kakimoto \cite{ogio}, as well as
	  Roulet et al. \cite{roulet}.
	  Lower right panel: Propagation models by
	  Dova et al. \cite{dova},
	  Lagutin et al. \cite{lagutin},
	  Swordy \cite{swordy}, and
	  Tkaczyk \cite{tkaczyk}.
	  In addition, the range of $\lna$ for the extrapolation of direct
	  measurements according to the \modell\ is indicated as dotted grey
	  lines, see text.}
 \label{lnamod}	  
\end{figure*}

Taking the experimental values from experiments measuring particle
distributions at ground level \cite{polygonato} and results from experiments
observing the average depth of the shower maximum according to an
interpretation with a modification of QGSJET (model~3a in \cite{wq}), the
average mean logarithmic mass has been calculated.  For details on the
experiments and references the reader is referred to \cite{polygonato,wq}. The
mean value and the variation as function of primary energy are given in
Table~\ref{massetab} for reference.  The average experimental values are also
presented graphically as shaded area in Figure~\ref{lnamod} together with the
predictions of various models as described below in section~\ref{theosect}.

The values calculated with the \modell\ are presented in Figure~\ref{lnamod} in
addition as grey lines.  The inclusion of ultra-heavy elements has only a minor
effect on $\lna$, at $10^8$~GeV the values with and without ultra-heavy
elements differ only by $\Delta\lna\approx0.2$.  The dotted lines depict the
systematic errors of the model resulting from the uncertainties of the direct
measurements and the errors of the fit parameters of the \modell.  In order to
include also systematic effects caused by the assumptions made in the model,
$\lna$ has been calculated for $1\le Z\le28$ and $1\le Z\le92$, with a common
$\gamma_c$ and a common $\Delta\gamma$, as well as with and without {\sl
ad-hoc} component. The largest and smallest values obtained with these
combinations are shown in the figure, representing the total systematic error.
The errors are asymmetric, since compared to heavy elements the energy spectra
of light elements are specified more precisely by direct measurement.

It should be mentioned that at energies below 1~PeV the indirect observations
obtain a lighter mass composition than the direct measurements. The $\lna$
range of the latter is bounded by the dotted grey lines.

\section{Theoretical models for the \knie} \label{theosect}
Proposed explanations in the literature for the \knie\ may be divided into four
categories. The first three discuss astrophysical reasons, arguing that the
\knie\ is an intrinsic property of the energy spectrum, while in the last class
the authors consider new particle physics processes in the atmosphere. In
these theories the \knie\ does not exist in the primary energy spectrum, but 
is only an effect of observing extensive air showers in the atmosphere:

1) Models in the first category relate the \knie\ to the acceleration process.
The standard approach of particle acceleration in shock fronts from supernova
explosions and several extensions are discussed in the models by Berezhko and
Ksenofontov~\cite{berezhko}, Stanev et al.~\cite{stanev}, Kobayakawa et al.
\cite{kobayakawa}, and Sveshnikova \cite{sveshnikova}.  The maximum energy
attained is taken to be responsible for the \knie.  Also source related is the
proposal of Erlykin and Wolfendale~\cite{wolfendale}, namely that a nearby
supernova as a single source of cosmic rays causes the structure in the energy
spectrum.  V\"olk and Zirakashvili \cite{voelk} consider reacceleration by
spiral shocks in the galactic wind.  Plaga \cite{plaga} proposes cosmic rays
being accelerated by cannon balls ejected into the galactic halo.

2) Models of the second category connect the \knie\ with leakage of cosmic rays
from the Galaxy.  Swordy~\cite{swordy} combines diffusive shock acceleration
with an energy dependent propagation pathlength in a Leaky Box model.  In the
models by Lagutin et al.~\cite{lagutin}, Ptuskin et al.~\cite{ptuskin}, Ogio
and Kakimoto \cite{ogio}, as well as Roulet et al. \cite{roulet} the \knie\
follows from diffusive propagation of cosmic-ray particles through the Galaxy. 

3) Interactions of cosmic rays with background particles in the Galaxy are
considered as origin for the \knie\ in the third category.
Tkaczyk~\cite{tkaczyk} suggests diffusive propagation in combination with
photo-disintegration.  Interactions of cosmic rays with background photons are
proposed by Dova et al.~\cite{dova}. Candia et al.~\cite{candia} discuss the
interaction of cosmic rays with the neutrino background.

4) A last class of theories accounts the air shower development in the
atmosphere for the \knie.  The basic idea is that a new type of interaction
transfers energy to a component not observable (or not yet observed) in air
shower experiments.  The threshold for these new interactions is in the \knie\
region, above the energy of todays collider experiments. In the model by
Kazanas and Nicolaidis the energy is transfered into techni-hadrons, lightest
supersymmetric particles \cite{kazanasall}, and gravitons \cite{kazanasgrav}.

In the original publications different representations for the theoretical
spectra are used, including different sets of mass groups and energy
multipliers for the flux values.  In order to provide an overview on the
predicted spectra and to allow an easy comparison, all results are presented in
a standard format with the ordinate being multiplied by $E^{2.5}$.  The energy
spectra are given for the following five mass groups: protons (Z=1), helium
(Z=2), intermediate: lithium -- fluorine (Z=3--9), heavy: neon -- titanium
(Z=10--22), and very heavy: vanadium -- uranium (Z=23--92).

To provide some orientation for the comparison of different models, in the
figures depicting the results of the individual models also calculations
according to the \modell\ are shown. As has been discussed in
section~\ref{pgsect} these spectra are reasonable approximations of the
spectra for elemental groups as derived from air shower measurements.

\subsection{Acceleration in supernova remnants}           
\begin{figure*}\centering
 \epsfig{file=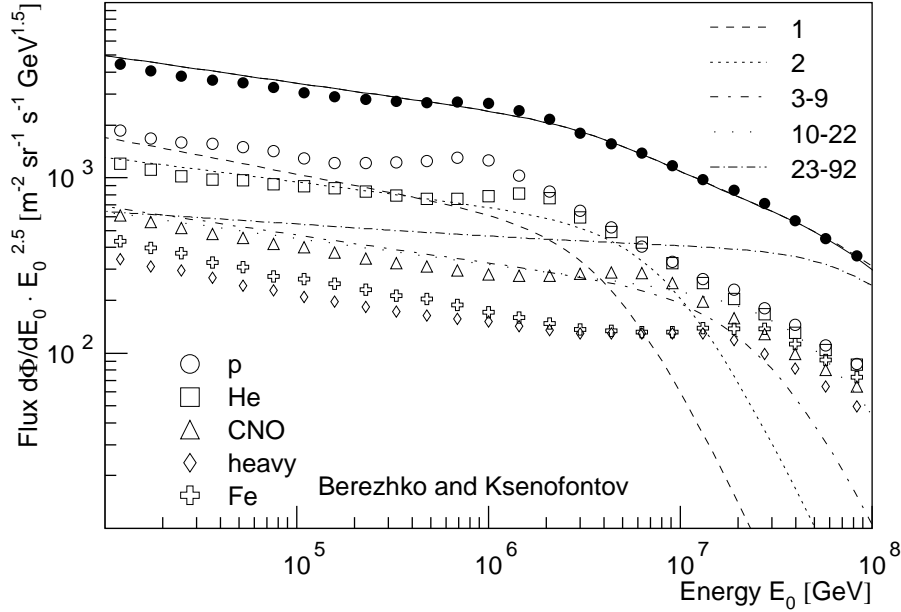,width=12cm}
 \caption{Energy spectra for five groups of elements according to 
	  the model by Berezhko \& Ksenofontov \cite{berezhko} (open symbols)
	  and the \modell\ (dashed and dotted lines), the numbers indicate the
	  range of nuclear charge for each group. Filled points and the full
	  line represent the all-particle spectra.}
 \label{berezhko}	  
\end{figure*}

The first four models deal with different variations of diffusive shock
acceleration of cosmic rays in shock fronts.  The maximum energy reached in
this process is related to the \knie\ in the energy spectrum.  The calculations
by Berezhko and Ksenofontov \cite{berezhko} are based on the nonlinear kinetic
theory of cosmic-ray acceleration in supernova remnants.  The mechanical energy
released in a supernova explosion is found in the kinetic energy of the
expanding shell of ejected matter. The cosmic-ray acceleration is a very
efficient process and more than 20\% of this energy is transfered to ionized
particles. The acceleration is achieved in the following way: The diffusive
propagation of high-energy particles allows them to cross the shock front many
times and every pair of consecutive crossings of the shock front increases the
energy of the particles.  In linear approximation this process generates a
power-law momentum spectrum at a planar shock front. Due to the high
acceleration efficiency and the hardness of the spectrum the structure of the
shock wave is modified by the reciprocal influence of cosmic rays on the medium
and, in consequence, the cosmic-ray spectrum is altered.

The particles need some minimum velocity in order to cross the shock front.
This velocity determines the injection rate of particles into the acceleration
regime. Superthermal particles can efficiently be injected and accelerated with
high efficiency. The injection efficiency is expected to be an increasing
function of the mass to charge ratio ($A$/$Z$) of the nucleus considered, i.e.
heavy elements are accelerated more efficiently and obtain a harder spectrum.
The energy limit of cosmic rays accelerated in supernova remnants is controlled
by geometrical factors of the expanding shock front.  Taking into account
preacceleration in the wind of the predecessor star maximum energies up to
$Z\cdot10^{15}$~eV can be achieved.  In regions modified by the presupernova
wind the magnetic fields are larger than in the unperturbed interstellar
medium. For the calculations the density of elements heavier than hydrogen in
the interstellar medium is assumed to be proportional to the relative abundance
of elements in the local, solar region of the Galaxy. The expansion of the
shock wave into a hot interstellar medium, which is assumed to be significantly
modified by presupernova winds of type Ib and type II supernovae, with a
magnetic field $B_0=12~\mu$G has been taken into account with an injection rate
$\eta=5\cdot10^{-4}$.

The resulting energy spectra for five elemental groups are plotted in
Figure~\ref{berezhko} (symbols). The \knie\ in the all-particle spectrum is due
to the charge dependence of the maximum energy achieved in the acceleration
process. The proton component exhibits a \knie\ at 1~PeV with a relatively
small change of the spectral index which keeps the protons to be the dominant
component up to $10^8$~GeV. Elemental spectra according to the \modell\ are
represented by the lines. They show a steeper decrease of the flux above the
\knie\ for all elements. Below 1~PeV the spectra for He and the CNO-group of
the model by Berezhko \& Ksenofontov agree with the \modell, but the heavy
elements are suppressed in favor of the protons.  On the other hand, the
all-particle spectra (solid line and filled dots) are well compatible in shape
and absolute normalization.

The mean logarithmic mass has been calculated using the five mass groups, the
result is shown in Figure~\ref{lnamod}  as function of energy\footnote{Only
$\langle A\rangle$ versus energy is given in the original article.}.  The
values exhibit a slight decrease up to about $10^6$~GeV to a minimum value
around $\lna =1.2$. At higher energies $\lna$ increases and reaches $\lna=2.1$
at $10^8$~GeV.  The rise of $\lna$ beyond the \knie\ is relatively modest for
the model in question, mainly caused by the moderate flux decrease beyond the
individual \knie s.  In the whole energy range the $\lna$ values are at the
lower edge or even below the range of measured values.

\subsection{Acceleration by supernova shocks} 
\begin{figure*}\centering
 \epsfig{file=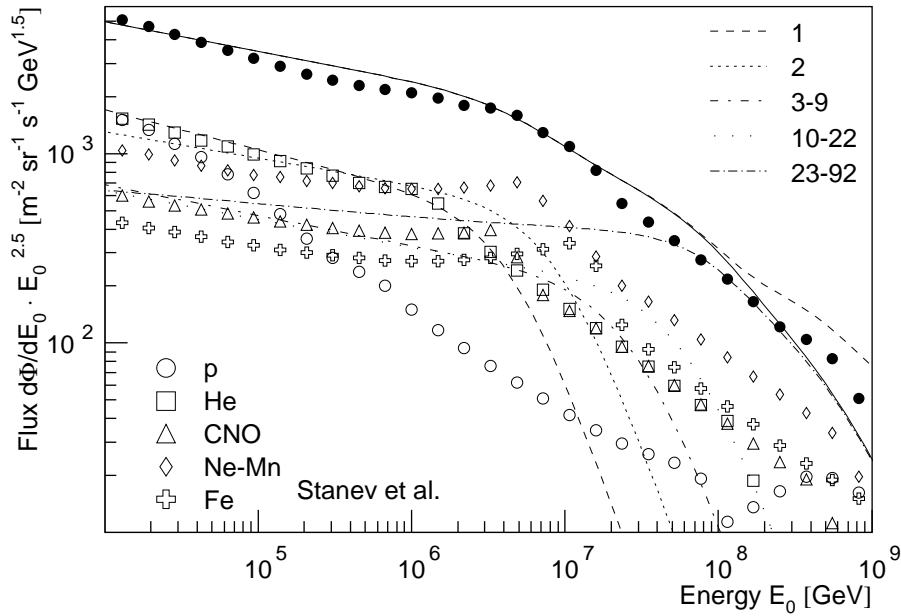,width=12cm}
 \caption{Energy spectra for five groups of elements according to
	  the model by Stanev, Biermann, and Gaisser \cite{stanev} (open
	  symbols) and the \modell\ (dashed and dotted lines).  The
	  numbers indicate the range of nuclear charge for the elemental
	  groups. Filled points show the sum spectrum for all groups according
	  to the theoretical calculations. The sum spectrum for all elements
	  according to the \modell\ is represented by a solid line, above
	  $10^8$~GeV a dashed line represents the average experimental flux
	  spectrum (see \cite{polygonato}).}
 \label{stanev}	  
\end{figure*}

A threefold origin of energetic cosmic rays is proposed by Stanev, Biermann,
and Gaisser \cite{stanev}. In their model particles are accelerated at three
different main sites \cite{biermann}: \\
1) The explosions of normal supernovae into an approximately homogeneous
interstellar medium drive blast waves which can accelerate protons to about
$10^5$ GeV. Particles are accelerated continously during the expansion of the
spherical shock-wave, with the highest particle energy reached at the end of
the Sedov phase.\\
2) Explosions of stars into their former stellar wind, like that of Wolf Rayet
stars, accelerate particles to higher energies.  The maximum energy attained
depends linearly on the magnetic field and maximum energies $E_{max}=9\cdot
10^7$~GeV for protons and $E_{max}=3\cdot 10^9$~GeV for iron nuclei are
reached.\\
3) For energies exceeding $10^8$~GeV an extragalactic component is introduced
by the authors. The hot spots of Fanaroff Riley class II radio galaxies are
assumed to produce particles with energies up to $10^{11}$~GeV.

In this model the main fraction of galactic cosmic rays above about 10~TeV are
assumed to be accelerated by shocks that travel down a steady stellar wind with
a Parker spiral structure for the magnetic field. Like in other first order
Fermi theories the basic idea for the energy gain is the cyclic crossing of the
shock wave from upstream to downstream and back. In most regions the shock
normal is perpendicular to the magnetic field. But in small regions around the
poles (about 1\% of $4\pi$~sr) the direction of the propagation of the shock is
parallel to the magnetic field which yields harder spectra for the accelerated
particles. The combination of the polar cap with the rest of the stellar
hemisphere might lead to a situation where up to about $10^4$~GeV the entire
hemisphere excluding the polar cap dominates, while from $10^4$~GeV up to the
\knie\ the polar cap begins to contribute appreciably.  At \knie\ energies the
polar cap component contributes equally to the remainder of the $4\pi$~sr solid
angle. This fact is essentially to describe a sharp bend in the all-particle
spectrum as it has been observed by some experiments, e.g. Akeno \cite{akeno}.
The logical consequence of the model is that the chemical composition at the
\knie\ changes in a way, that the gyroradii of the particles at the spectral
break are identical. This implies that the fluxes of different nuclei break
off according to their charge $Z$.

Taking the chemical composition as measured directly by experiments up to
100~TeV energies, Stanev et al. fit the shower size spectra obtained by the
Akeno experiment for different zenith angles and derive primary energy spectra
for groups of elements. The spectra obtained for five mass groups are
reproduced in Figure~\ref{stanev}. The proton flux starts to deviate from a
power law at $10^4$~GeV caused by the energy limit reached for the Sedov phase
expansions in the interstellar medium and the flux decreases up to $10^8$~GeV.
The increase of the proton flux above this energy is caused by the
extragalactic component introduced. Elements heavier than hydrogen exhibit a
sharp, rigidity dependent bend.  

\begin{figure*}\centering
 \epsfig{file=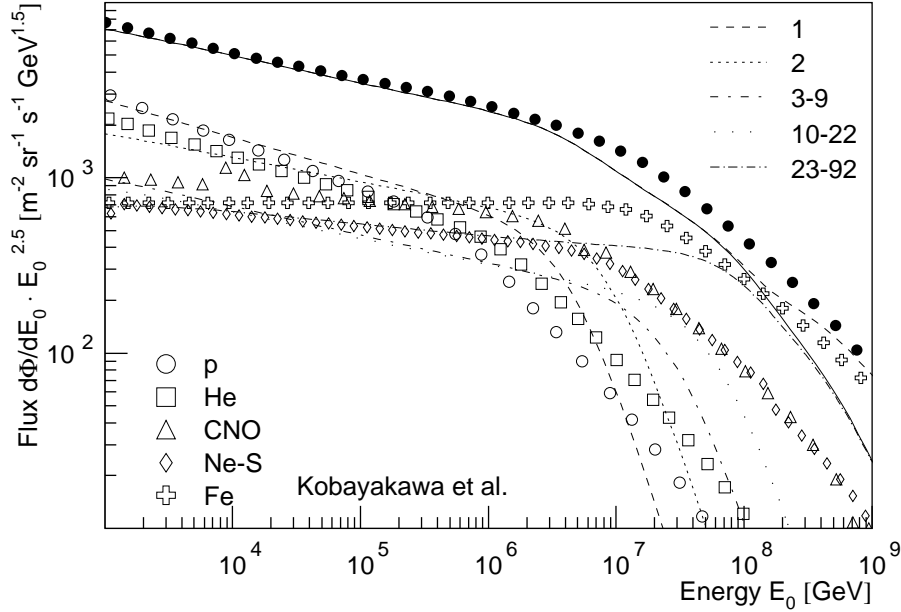,width=12cm}
 \caption{Energy spectra for groups of elements for the model by Kobayakawa 
	  et al. \cite{kobayakawa} (symbols) and the \modell\ (lines),
	  see caption Figure~\ref{stanev}.}
 \label{koba}	  
\end{figure*}

The relative distinct structure of the \knie\ is due to the contribution of the
polar cap component. Beyond the \knie\ the calculated spectra are flatter as
compared to the \modell\ and the contribution of light elements is larger. The
sharp \knie s for the individual element spectra result in a pronounced
structure in the all-particle spectrum.  The model has been tuned to describe
the sharp \knie\ seen by early experiments like MSU and Akeno. Recent
measurements yield a smoother transition at the \knie, see e.g. Figure~15 in
\cite{polygonato}. A fit to the more recent data would probably result in a
smaller contribution of the polar cap component.

The mean logarithmic mass calculated from the individual element spectra, shown
in Figure~\ref{lnamod}, exhibits an increase up to $\lna\approx3.1$ at an
energy $E_0\approx10^7$~GeV.  The early increase is caused by the early
decrease of the proton flux mentioned.  The $\lna$ values predicted by the
model increase faster as function of energy than the observations, but reach
their maximum at lower energies and smaller $\lna$ values are predicted at the
maximum.  The first peak is due to the \knie s in the spectra for the heavy
elements (Ne and Fe groups) at energies around $10^7$~GeV.  The second peak and
the decreasing $\lna$ values are caused by the extragalactic proton component.

\subsection{Acceleration by oblique shocks}   \label{seckoba}
A slightly modified version of the diffusive acceleration of particles in
supernova remnants is considered by Kobayakawa, Sato, and Samura
\cite{kobayakawa}.  Standard first order Fermi acceleration in supernova
remnants --- with the shock normal being perpendicular to the magnetic field
lines --- is extended for magnetic fields with arbitrary angles to the velocity
of the shock front. The basic idea is that particles are accelerated to larger
energies in oblique shocks as compared to parallel shocks.

The ejected material from a supernova explosion expands into the interstellar
medium driving a shock wave. The shock acceleration may be effective until the
expanding shell sweeps up its own mass of the interstellar medium. The ejected
mass ($\approx10M_\odot$) with large momenta moves out freely at constant
velocity during the free expansion phase which is assumed to be in the order of
a few hundred years for ejecta expanding at a velocity $\approx 10^8$~cm/s into
a medium of an average density $\approx1$~proton/cm$^3$. The shape of the shock
front is supposed to be almost spherically symmetric. The directions of the
interstellar magnetic field lines will be, over a wide range, nearly at random
rather than well aligned. Therefore, it is assumed that the field lines meet
the shock front at random angles and the cosines of the angles are distributed
uniformly.  

The injection efficiency into the acceleration regime is investigated as
function of the angle between the magnetic fields and the normal of the shock
front. It is found that the injection efficiency decreases drastically as the
shock obliquity increases.  The simple relation for the efficiency
$\epsilon(\eta)=\eta$ is assumed, $\eta$ being the cosine of the angle between
the magnetic field and the expansion direction of the shock front.  On the
other hand, the obliquity makes the energy spectrum harder, particles are
accelerated more efficiently.  

For oblique shocks, reflection at the shock front is more important than
diffusion in downstream regions, so that a more rapid acceleration can occur
than at parallel shocks.  The maximum energies of accelerated particles can be
two or even three orders of magnitude larger in the case of a
quasi-perpendicular shock than in a parallel one.  The maximum energy per
particle reached during the acceleration process is mainly limited by the
finite lifetime of the supernova blast wave.  In the article a rigidity 
dependent maximum energy 
\begin{equation}
 \begin{array}{ll}
   \textstyle E_{max}\propto & 
   \textstyle Z \cdot 2.5\cdot10^{16}~\mbox{eV} \\
  & \textstyle
  \cdot \frac{\textstyle B}{\textstyle 30~\mu\mbox{G}} 
  \cdot \frac{\textstyle u}{\textstyle 10^7~\mbox{m/s}} \cdot f(\eta) \\
 \end{array}
 \label{eqkoba}
\end{equation}
is derived, with the speed of the shock wave $u$, the upstream magnetic field
$B$ --- being about ten times larger than conventionally estimated ---, and
$f(\eta)$ being an angle dependent acceleration efficiency.  For parallel,
strong shocks ($\eta=1$) a maximum energy $E_{crit}=Z\cdot1.25\cdot10^{14}$~eV
is obtained. This value is exceeded by more than a factor of 1400 for shocks
with extreme obliquity.

The authors fit experimental data for several groups of elements. The spectral
indices of the power laws as well as the absolute flux are adjusted at 1~TeV.
The resulting spectra are extrapolated to high energies as presented in
Figure~\ref{koba}. The individual \knie s result from the dependence of the
maximum energy reached on the angle between the shock and the magnetic field.
The spectra exhibit a rigidity dependent \knie.  A comparison with the \modell\
exhibits some differences, but the overall agreement is quite good.  However,
above the \knie s the model yields slopes less steep and overestimates the
all-particle flux.

The cut-off behavior of the iron-group is the reason for the strong increase
of the mean logarithmic mass as function of energy, as shown in
Figure~\ref{lnamod}. The model predictions exhibit an increase up to
$\lna\approx3.5$ at $10^8$~GeV, very close to the maximum of the \modell.  The
$\lna$ values are at the upper border or even above the experimental values in
the whole energy range.

\subsection{Acceleration by a variety of supernovae}

In a recent article Sveshnikova revised the standard approach of acceleration
of cosmic rays in shock fronts of supernovae \cite{sveshnikova}.  The new
approach to understand acceleration in supernova remnants (SNRs) is based on
recent astronomical observations of supernovae \cite{hamuy,turatto}. They
indicate that core collapse supernovae prove to comprise the most common
general class of exploding stars in the universe and they occur in a great
variety of flavors.  In particular, supernovae have been observed with unusual
large expansion velocities, indicating that these objects are hyper-energetic,
the so called hypernovae.

The upper limit of acceleration in SNR is determined essentially by the product
of the shock radius $R_{sk}$, shock velocity $V_{sk}=v_{sk}/1000~{\rm km/s}$,
ejected mass $M_{ej}$, remnant age $T_{snr}$, and explosion energy
$E_{51}=E_{snr}/10^{51}~{\rm erg}$.  All these values are connected to each
other and vary from explosion to explosion. The cut-off energy per particle
$E_{max} [{\rm TeV}]$ can be expressed by a simple formula if only the Sedov
Phase of SNR expansion is considered \cite{ellison}  
\begin{equation}
\begin{array}{rl}
E_{max} =
  & 200 \cdot Z \left(\frac{0.3 \cdot B}{3~\mu{\rm G}}\right)
              \left(\frac{n_H}{{\rm cm}^3}\right)^{-\frac{1}{3}} \\
  & \cdot \left(\frac{E_{snr}}{10^{51}~{\rm erg}}\right)^\frac{1}{3}
    \left(\frac{v_{sk}}{10^3~{\rm km/s}}\right)^\frac{1}{3}\\
= & Z\cdot E_{max}^0(B,n_H)\left(E_{51}\cdot V_{sk}\right)^\frac{1}{3}\\
\end{array}
\end{equation}
The maximum energy depends on three factors, the charge $Z$ of the nucleus, the
properties of the interstellar medium where the SNR is expanding (strength of
magnetic field $B$ and density of protons $n_H$), and on the energy of the
explosion as well as the velocity of the shock.  

\begin{figure*}\centering
 \epsfig{file=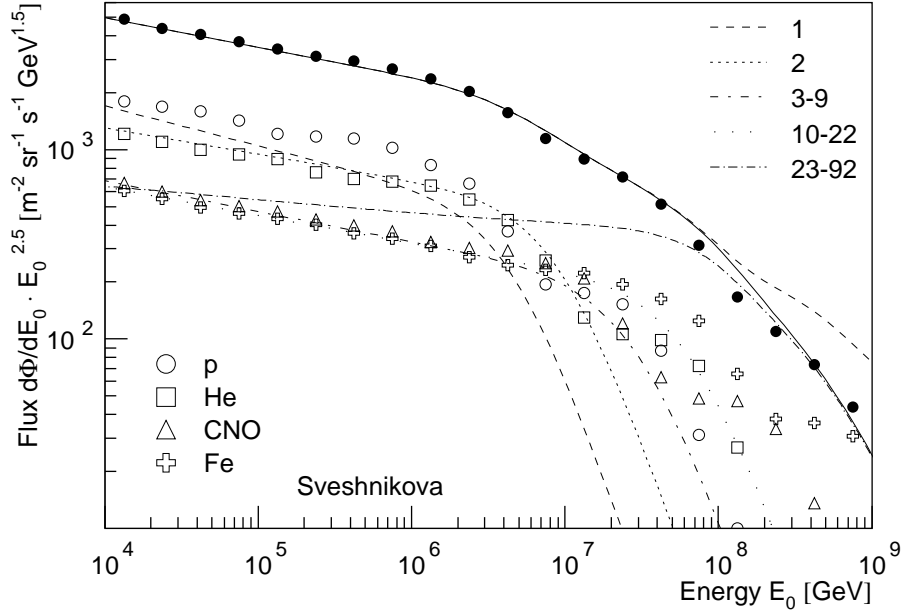,width=12cm}
 \caption{Energy spectra for groups of elements for the model by 
	  Sveshnikova \cite{sveshnikova} (open symbols) and the \modell, see
	  caption of Figure~\ref{stanev}.}
 \label{sveshnikova}	  
\end{figure*}

A usual way to increase $E_{max}$ is to enlarge the magnetic field $B$, see
e.g. the approach by Kobayakawa et al. (section \ref{seckoba}).  On the other
hand, as consequence of the recent observations also the parameters of the supernova
explosion can be varied. This is the basic idea of Sveshnikova, to draw up a
scenario in which the maximum energy reached in SNR acceleration is in the
\knie-region of the cosmic-ray spectrum ($\approx 4$~PeV), using only the
standard model of cosmic-ray acceleration and the latest data on supernovae
explosions.

Based on recent observations the distribution of explosion energies $E_{51}$
and their rates of occurrence in the galaxy are estimated.  The spectrum of
cosmic rays in each explosion is approximated by a composite power law
\begin{equation}\begin{array}{l}
\Phi(E)=\Phi_0 E^{-\gamma}\quad \mbox{with} \\
    \gamma=\left\{ 
    \begin{array}{ll}
     2.0 &; 10~{\rm GeV} < E < E_{max}/5\\
     1.7 &; E_{max}/5 < E < E_{max}\\
     5.0 &;  E > E_{max}/5 \quad.\\
    \end{array}\right.\\
    \end{array}
\end{equation}

The observed spectrum in the Galaxy is obtained as sum over all different types
of supernovae explosions, integrated over the distribution of explosion
energies within each supernova group.  The mass composition of cosmic rays has
been assumed according to direct measurements.  As result energy spectra at the
source are obtained for groups of elements.  In order to take into account
propagation effects the spectra given in \cite{sveshnikova} have been
multiplied with $E_0^{-0.65}$, as suggested in the article. The flux has been
normalized to the average measured all-particle spectrum at low energies and
the resulting spectra are depicted in Figure~\ref{sveshnikova}.  

Since the spectra for elemental groups are the result of many supernovae
explosions, they can not be described by a simple function with two power laws.
Instead, they exhibit a more complicated structure, the cut-off above the
individual \knie s is characterized by two subsequent steps, caused by
contributions of individual supernovae classes.  The the first step for each
elemental group occurs at about the same energy as in the \modell. At energies
larger than the individual cut-offs, the spectra predicted by Sveshnikova are
significantly flatter due to the contribution of very energetic SNRs.

\begin{figure*}\centering
 \epsfig{file=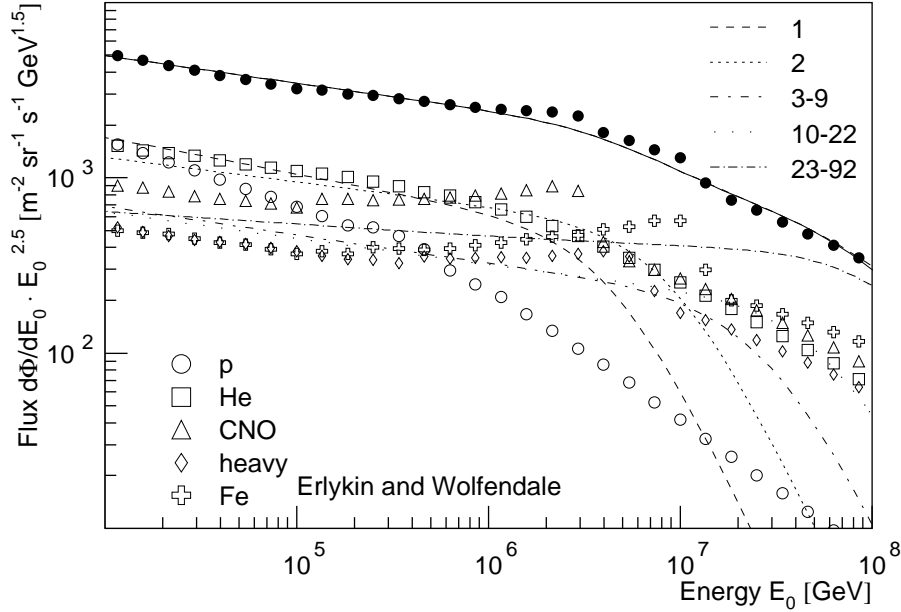,width=12cm}
 \caption{Energy spectra for groups of elements 
	  for the model by Erlykin \& Wolfendale \cite{wolfendale} (open
	  symbols) and the \modell, see caption of
	  Figure~\ref{stanev}.}
 \label{wolfendale}	  
\end{figure*}

The \knie\ in the all-particle spectrum is caused by the cut-off of the proton
component. The all-particle spectrum above the \knie\ is formed by the sum of
coupled steps. The change of the slope beyond the \knie\ (in the interval from
3~PeV to $Z_{Fe}\cdot 3$~PeV) is determined by the fraction of iron nuclei in
the chemical composition of cosmic rays before the \knie. The maximal energy of
Galactic cosmic rays is determined by iron nuclei generated in the most
energetic supernovae (type SNIIn) and corresponds to about $10^9$~GeV.

The iron component is clearly dominant at high energies, which leads to a
strong increase of the mean logarithmic mass, as can be inferred from
Figure~\ref{lnamod}. A slight structure is visible around 10~PeV, at this
energy the light nuclei (protons and helium) are replaced by heavy nuclei as
the major component in cosmic rays. The mean logarithmic mass increases to
values above $\lna=3.5$ at $10^9$~GeV. Additionally, a variant is discussed,
where cosmic rays from SNIIn are enriched by heavy nuclei. This results in a
heavier mass composition. In the region between $10^7$ and $10^8$~GeV $\lna$ is
increased up to 1 unit. The model predictions for both variants are well
compatible with the average experimental values. The latter scenario exhibits
good agreement with the \modell.

\subsection{The single-source model}   

A different approach is used by Erlykin and Wolfendale \cite{wolfendale}. In
addition to a background caused by many undefined sources the authors presume a
single supernova remnant from a recent and nearby explosion as additional
source of cosmic rays. The \knie\ in the all-particle spectrum is supposed to
have a two-kink structure related to the cut-offs of oxygen and iron nuclei
from the single source.  The authors use shower size spectra of the
electromagnetic, muonic, hadronic and \v{C}erenkov components of extensive air
showers. The shower size spectra are normalized to the same \knie\ position.
The proposed twofold structure is not visible in all individual shower size
spectra considered, it shows up after the normalization and rebinning
procedure, combining results from several experiments.

In the publication energy spectra for groups of elements are given for both,
the background and the single SNR separately. The sum of both components is
shown in Figure~\ref{wolfendale} for individual element groups. Salient kinks
in the spectra of the CNO and Fe groups caused by the single SNR produce\ a
twofold structure in the all-particle spectrum at $3\cdot 10^6$~GeV and
$10^7$~GeV. Such a structure is not observable in the averaged all-particle
spectrum of the measurements. In the single-source model the {\knie}s of the
elements heavier than helium are much sharper than in the \modell.

The mean logarithmic mass taken from Erlykin and Wolfendale is shown in
Figure~\ref{lnamod}. The early decrease of the proton component results in a
strong increase of $\lna$ at the upper border of the observed values.  Within
the present experimental resolution, the structure at $10^7$~GeV can not be
resolved.

\subsection{Recceleration in the galactic wind}

\begin{figure*}\centering
 \epsfig{file=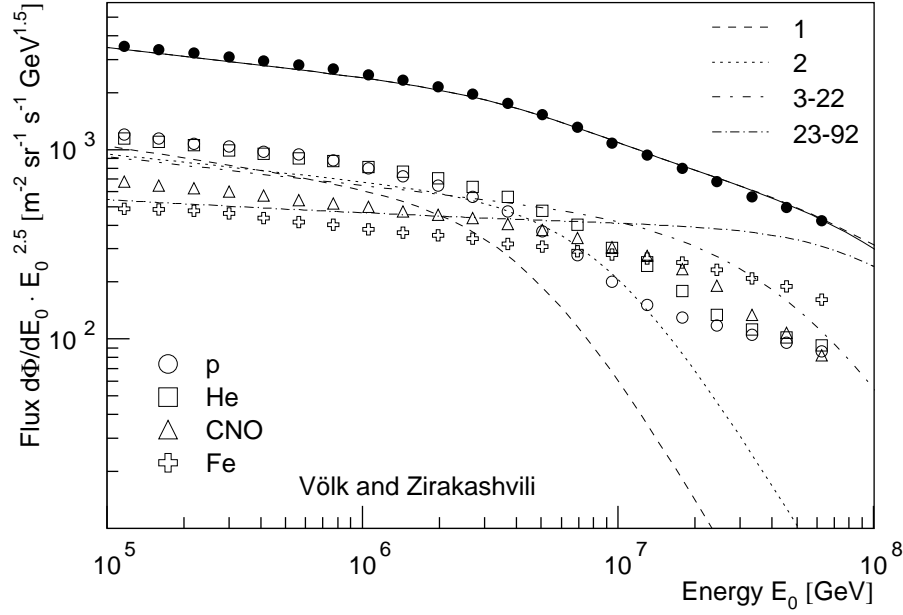,width=12cm}
 \caption{Energy spectra for groups of elements 
	  for the model by V\"olk and Zirakashvili \cite{voelk} (open symbols)
	  and the \modell, see caption of Figure~\ref{stanev}.}
 \label{voelk}
\end{figure*}

Reacceleration of cosmic-ray particles in the Galactic wind is discussed by
V\"olk and Zirakashvili \cite{voelk}.  The wind is mainly driven by cosmic rays
and hot gas generated in the disk.  It reaches supersonic speeds at about
20~kpc above the disk, and is assumed to be very extended (several 100~kpc)
before it ends in a termination shock.  Due to Galactic rotation the
differences in flow speed will lead to strong internal wind compressions,
bounded by smooth cosmic-ray shocks. These shocks are assumed to reaccelerate
the most energetic particles from the disk by about two orders of magnitude in
rigidity, ensuring a continuation of the energy spectrum beyond the \knie\ up
to the ankle. A fraction of the reaccelerated particles will return to the
disk, filling a region around the Galactic plane (several tens of kpc thick)
rather uniformly and isotropically. 

A maximum energy $E_{max}\approx Z\cdot10^{17}$~eV is obtained and the authors
conclude that the \knie\ in the all-particle spectrum cannot be the result of
the propagation process, instead it is supposed to be a feature of the source
spectrum itself. It is pointed out that it is possible to explain the
continuation of the cosmic-ray spectrum above the \knie\ up to the ankle in a
natural way, by considering the dynamics of the interstellar medium of the
Galaxy and its selfconsistent extension into a large-scale halo by the Galactic
wind. The authors conclude further that within this picture there is no way to
produce higher energy cosmic rays, their sources must be of a different nature.

\begin{figure*}\centering
 \epsfig{file=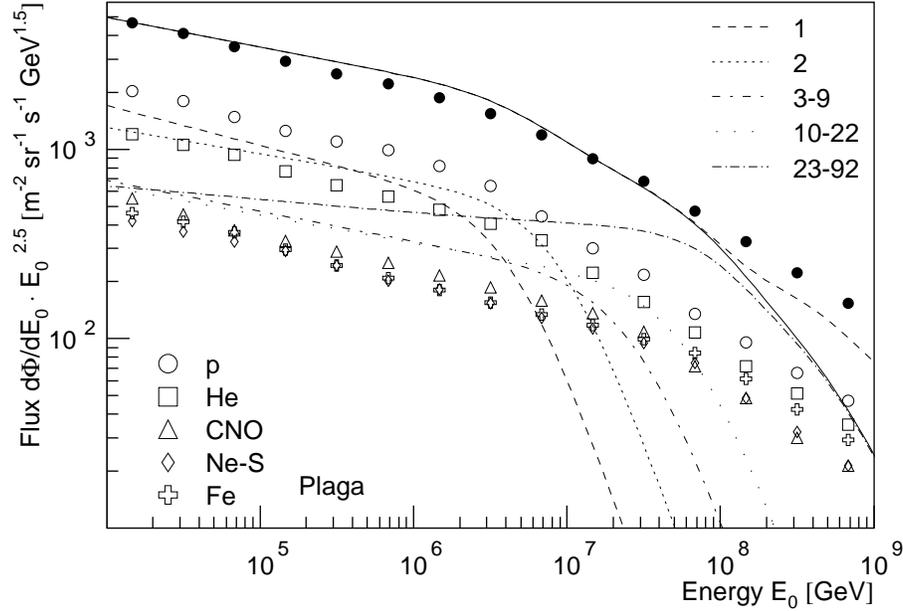,width=12cm}
 \caption{Energy spectra for groups of elements for the model by 
	  Plaga \cite{plaga} (open symbols) and the \modell, see caption of
	  Figure~\ref{stanev}.}
 \label{plaga}
\end{figure*}

For the calculations the chemical composition of cosmic rays has been adjusted
at $9\cdot10^{14}$~eV to KASCADE measurements. The energy spectra derived for
four elemental groups as well as the all-particle spectrum are depicted in
Figure~\ref{voelk}.  For the light elements (protons and helium) a \knie-like
structure is clearly visible.  For these elemental groups the decrease at high
energies is significantly attenuated and the spectra continue at high energies
with a spectral slope similar to the values below the \knie. This behavior
differs from the \modell\ with its assumptions on the cut-off behavior,
while the all-particle spectra agree well with each other. 

The mean logarithmic mass deduced is depicted in Figure~\ref{lnamod}.  In the
model iron nuclei are only about a factor of two more abundant than protons at
high energies ($E_0\approx10^8$~eV). Consequently, only a modest increase of
the mean logarithmic mass is obtained with a maximum value
$\lna_{max}\approx2.6$.  The predictions are well compatible with the average
experimental $\lna$ values below $10^7$~GeV, but may be slightly too low at the
highest energies.

\subsection{The cannonball model}

Within the framework of the cannonball model, originally proposed by Dar and De
R\'ujula \cite{dar} to explain gamma-ray bursts, Plaga \cite{plaga} discusses
a mechanism for the acceleration of cosmic-ray hadrons.  The author
investigates if masses of baryonic plasma ("cannonballs"), ejected in bipolar
supernova explosions, could be the universal sources of hadronic galactic
cosmic rays.  It is assumed, that in each of the symmetric cannonball jets a
total energy of $10^{53}$~erg is ejected. Supernova explosions which eject
cannonballs should occur in the Galaxy at rates of 1/(50~a) to 1/(600~a).  Less
than 1/40 of the total cannonball-energy has to be converted to the energy of
the cosmic-ray particles in order to match the observed cosmic-ray
luminosity.

Two scenarios for the acceleration are sketched.  The first is based on
ultra-relativistic shocks in the interstellar medium and could accelerate cosmic
rays up to \knie\ energies.  The second is based on second order Fermi
acceleration inside the cannonballs.  For cosmic-ray propagation a diffusion
coefficient $D\propto E^{-\alpha}$, with $\alpha=0.5$, is adopted. No
discontinuities in the energy dependence of the diffusion coefficient are
assumed, thus the \knie\ in the energy spectrum is related to source
properties.

The bulk of cosmic rays up to the \knie\ is taken to be accelerated at
ultra-relativistic shocks driven by cannonballs into the interstellar medium.
The accelerated particles are immediately swept up into the plasmoid and
remain confined there until it has slowed down to subrelativistic speeds in the
galactic halo.  The maximum energy 
\begin{equation}
 E_{max} = 3\cdot10^{15}~\mbox{eV} ~ \frac{R}{0.02~\mbox{pc}}
           \frac{Z\cdot B}{\mu\mbox{G}} \frac{\Gamma}{50}
\end{equation}
that can be reached in this acceleration process is proportional to the radius
of the plasmoid $R$, to its Lorentz factor $\Gamma$, and to the interstellar
magnetic field $B$. $Z$ is the nuclear charge of the accelerated particle.  A
fraction of particles contained inside the plasmoid escapes by diffusive
processes and determines the observed cosmic-ray luminosity.  A power law with
an spectral index $\gamma=-2.2$ at the source is obtained below the \knie.

\begin{figure*}\centering
 \epsfig{file=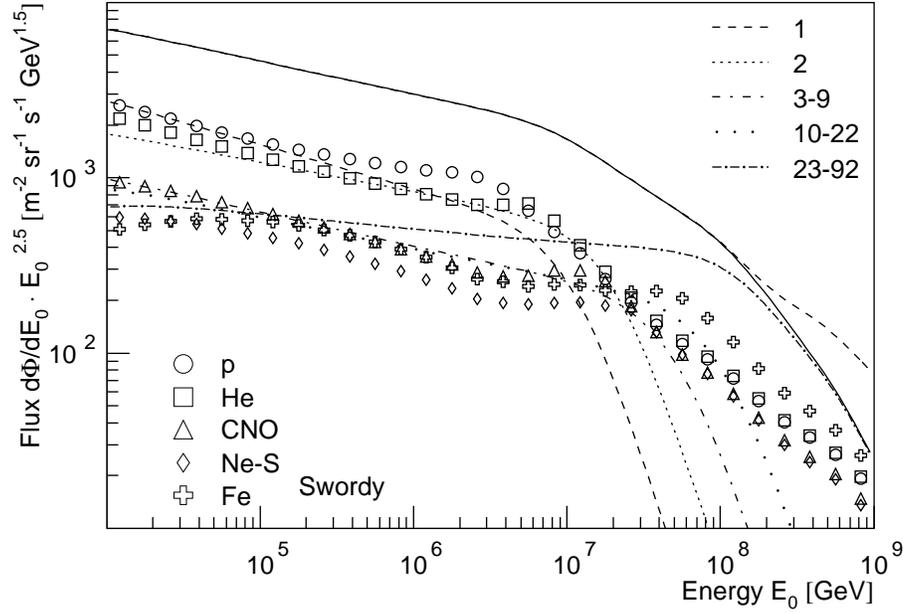,width=12cm}
 \caption{Energy spectra for groups of elements 
	  for the model by Swordy \cite{swordy} (open symbols) and the
	  \modell, see caption of Figure~\ref{stanev}.}
 \label{swordy}
\end{figure*}

Even higher energies could be reached inside the cannonballs by further
acceleration of contained hadrons in the internal turbulent plasma via second
order Fermi mechanisms. Typical acceleration times are in the order of several
$10^2$ to $10^5$~a and energies up to $Z\cdot10^{20}$~eV could be reached.
Again, the particles are diffusively released, their energy spectrum
follows a power law with an index $\gamma=-2.5$ at the source.

The energy spectra for groups of elements as obtained by Plaga are presented in
Figure~\ref{plaga}. The individual spectra exhibit a \knie\ proportional to the
charge. However, the change in spectral slope is extremely soft.  The
individual groups decline in the figure almost parallel, without any
significant change of the mass composition.  The obtained all-particle spectrum
is in reasonable agreement with the average measured flux.

In the whole energy range shown, light particles dominate the mass composition.
Therefore, only a small increase of the mean logarithmic mass is found above
the \knie, as can be inferred from Figure~\ref{lnamod}.  At an energy of
$8\cdot10^7$~GeV the maximum value reached is still below 2.  The extreme light
mass composition predicted by the model seems not to be compatible with the
experimental values.

\subsection{The minimum-pathlength model}

The second class of models discussed describes the propagation of cosmic rays
through the interstellar medium. In these scenarios the \knie\ is a consequence
of leakage of particles from the Galaxy.  The calculations by Swordy
\cite{swordy} are based on a Leaky Box model for the cosmic-ray propagation.
The spectra of particles accelerated by diffusive shocks are assumed to have
the same spectral slope $\gamma=-2.15$ for all elements at the source with a
rigidity dependent cut-off at a rigidity $R~\approx 10^{15}$~V.  Above this
cut-off the spectra are assumed to decrease as $dN/dR\propto R^{-3}$.  The
pathlength $\lambda_e$ for escape from the Galaxy declines with rigidity but
has some minimum value.  As function of $R$ it has been parametrized using the
expression
\begin{equation}
 \lambda_e=\lambda_0 \left(\frac{R}{R_0}\right)^{-\delta}+\lambda_r 
\end{equation}
with $\lambda_0=6$~g/cm$^2$, $\delta=0.6$, and $R_0=10$~GV. The value for the
minimum pathlength $\lambda_r=0.013$~g/cm$^2$ has been determined by a fit to
the measured all-particle spectrum.

Using this set of parameters, the author gives fractional abundances for five
elemental groups as function of energy.  They are normalized to the mean values
of direct measurements in the TeV energy range.  Taking the fractional
abundances and using the average all-particle spectrum as obtained from many
experiments, the energy spectra shown in Figure~\ref{swordy} are calculated.
The change of slope for the individual spectra is very smooth in the model by
Swordy. Above the \knie\ the spectra are significantly flatter as compared to
the \modell. The cut-off with $R^{-3}$ seems to be to smooth.

The individual spectra show a slight dip just below the individual {\knie}s.
This feature causes the dip in the mean logarithmic mass shown in
Figure~\ref{lnamod}. The $\lna$ values have been calculated from the individual
spectra shown in Figure~\ref{swordy}\footnote{The original publication gives
only values for the mean mass.}. They reach a maximum of $\lna\approx2.5$. The
modest increase obtained is due to the relative flat spectra above the
individual cut-offs, which yields to a mass composition, which is mostly at the
lower edge of the experimental values.

\subsection{Anomalous diffusion in the Galaxy}  
\begin{figure*}\centering
 \epsfig{file=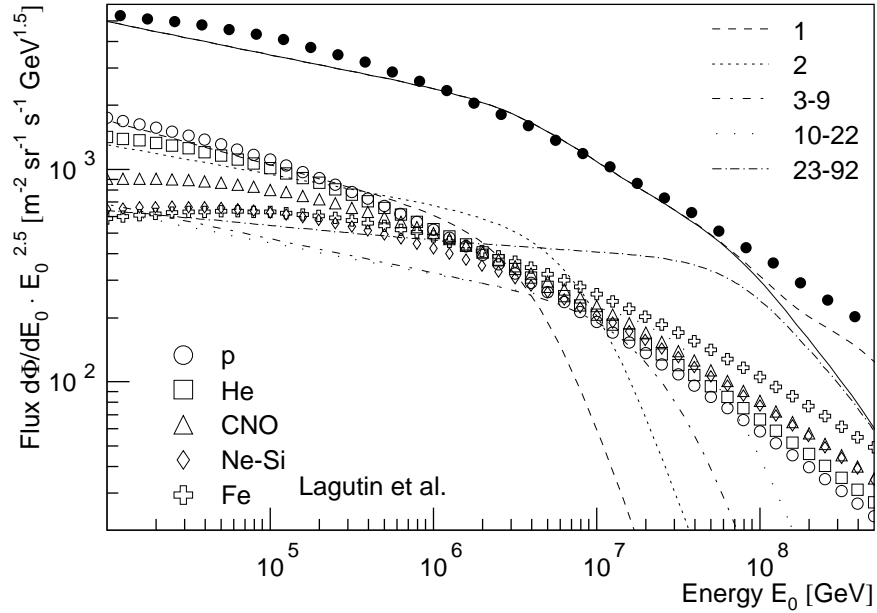,width=12cm}
 \caption{Energy spectra for groups of elements
	  for the model by Lagutin et al. \cite{lagutin} (open symbols)
	  and the \modell, see caption of Figure~\ref{stanev}.}
 \label{lagutin}
\end{figure*}

Lagutin et al. \cite{lagutin} discuss the \knie\ structure as being linked to
anomalous diffusion of cosmic rays in the magnetic fields of the Galaxy.  In
the literature several models are considered to describe the propagation of
cosmic rays through the interstellar medium. Frequently, a conception is used
in which the propagation process is determined by scattering at magnetic-field
inhomogeneities which have small-scale character. They can be considered as a
homogeneous Poisson ensemble and the propagation is described by a normal
diffusion model, e.g. as discussed by Berezinsky et al. \cite{berezinsky}. 

Lagutin et al. propose the existence of multiscale structures in the Galaxy,
related to filaments, shells, and clouds which are widely spread in the
interstellar matter. A rich variety of structures that are created in
interacting phases having different properties can be related to the
fundamental property of turbulence. The tool of fractal geometry is used to
characterize the interstellar medium and, correspondingly, the magnetic field
strength.  In such a fractal-like interstellar medium the cosmic-ray
propagation can not be described by normal diffusion.  Instead, an anomalous
diffusion model or superdiffusion is proposed.  The anomaly results from large
free paths (L\'evy flights) of particles between magnetic domains. These paths
are distributed according to an inverse power law.  Also rest-states of motion
in a trap (L\'evy waiting times) are investigated.

The diffusion coefficient is assumed to depend on the rigidity of the particles
as $D\propto D_0R^\delta$.  $D_0$ describes the propagation taking into account
both, long free paths and the probability to stay in magnetic traps during the
propagation. The anomalous diffusion coefficient is evaluated using
experimental data on the anisotropy of the cosmic-ray flux in the energy
region from $10^3$ to $10^4$ GeV/nucleus. In the model the spectral index as
function of primary energy is derived. The differential flux of all particles
is supposed to consist of two components. The first one describes the
contribution of nearby sources ($d<1$~kpc), including 16 supernova remnants,
being responsible for the high energy region including the \knie. The second
deals with the bulk of observed cosmic rays with energies from 0.1~GeV to
10~GeV from numerous distant sources ($d>1$~kpc).

The authors conclude that the \knie\ in the primary cosmic-ray spectrum and the
observed distinction in the spectral index of protons and other nuclei can be
explained by superdiffusion propagation.  The energy dependence of particle
spectra supports the hypothesis, that nearby supernova bursts are the source of
high-energy cosmic rays.

The calculated spectra are shown in Figure~\ref{lagutin}.  Between 1 and 50~PeV
absolute flux and shape of the all-particle spectrum are compatible with the
measured spectrum, but below and above these energies the calculations
overestimate the experimental all-particle flux.  The energy spectra for
individual elements exhibit a very smooth behavior in the \knie\ region, no
kink in the spectra is visible and no distinct energy for the \knie\ can be
specified.  At low energies the fluxes of the individual element spectra do not
agree with direct measurements and their extrapolation using power laws.

The model is characterized by a very smooth change of the spectral slopes.  At
the highest energies above $10^8$~GeV the iron-group is only about a factor of
two more abundant than protons.  The resulting mean logarithmic mass, presented
in  Figure~\ref{lnamod}, shows a slow --- in the logarithmic plot almost linear
--- increase as function of energy. At 400~PeV a value of $\lna\approx2.6$ is
reached.

\subsection{Drift in the global regular magnetic field of the Galaxy}  
\label{seckalmykov}
\begin{figure*}\centering
 \epsfig{file=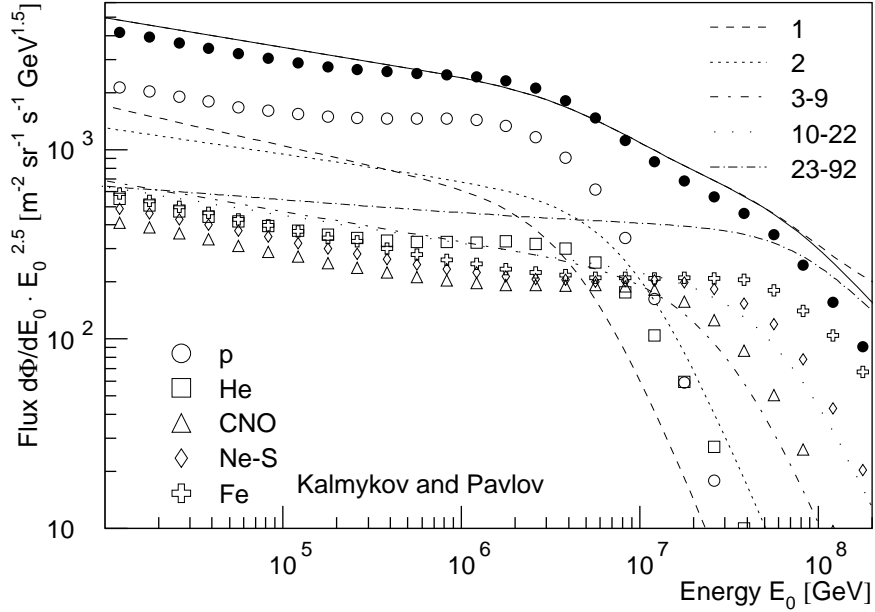,width=12cm}
 \caption{Energy spectra for groups of elements for the model by 
	  Ptuskin et al. \cite{ptuskin} adapted to experimental data by
	  Kalmykov and Pavlov \cite{kalmykov} (symbols) and the \modell, see
	  caption of Figure~\ref{stanev}.}
 \label{kalmykov}	  
\end{figure*}

The next three models described (sections~\ref{seckalmykov} to \ref{secroulet})
deal with variations of similar basic ideas. They treat the diffusion of cosmic
rays in the Galaxy taking into account the regular magnetic field, a random
component, and antisymmetric diffusion.

Ptuskin et al. \cite{ptuskin} demonstrate that a drift (hall diffusion) of
particles in the global regular magnetic field of the Galaxy may be responsible
for the \knie.  In the galactic halo a random field is assumed to exist
simultaneously with the regular magnetic field, which is mainly toroidal and
whose value is comparable with the value of the field in the gas disk. Particle
diffusion in a magnetically-active medium is described by a tensor.  The Hall
diffusion, viz. the antisymmetrical part of the cosmic-ray diffusion tensor, is
not effective at small energies, but it may control the leakage of cosmic rays
out of the Galaxy at energies above 3~PeV.  In this way the \knie\ in the
cosmic-ray power law spectrum is obtained without resorting to any assumptions
concerning this feature in the source spectrum or changes in the energy
dependence of the diffusion coefficient along the field lines.

The global structure of the regular magnetic field in the Galaxy is very
important to describe the propagation of cosmic rays since it determines the
orientation of the diffusion tensor. Reliable information on the regular field
in the disk is available, revealing that it is predominantly toroidal.
Observations indicate that the field lines of the regular field are closed
circles and the field changes sign with radial distance after each 3~kpc. It
has the same direction above and below the galactic plane.

Only scant information on the magnetic field beyond the galactic disk --- i.e.
at distances $>500$~pc --- exists. The orientation of the regular field in the
Galactic halo is not known, but the toroidal components seem to dominate.
Ptuskin et al. propose two scenarios, a flat halo model --- it is assumed that
the propagation region of cosmic rays in the Galaxy has the form of a cylinder
and the height of the cylinder is much smaller than its radius --- and a large
halo model. The latter is supported by radio astronomical data which indicate
that our Galaxy has an extended halo with a height of about 10~kpc comparable
to the galactic radius $\approx20$~kpc. Numerical calculations of the
cosmic-ray propagation in such an environment, assuming a power-law energy
spectrum with a constant spectral slope at the source, yield a cosmic ray
energy spectrum with a \knie\ at 3~PeV. 

At energies in the GeV region the Hall diffusion is insignificant and plays a
minor part in cosmic-ray leakage from the Galaxy.  The Hall diffusion
coefficient increases rapidly with energy and begins to dominate the
slowly-increasing usual diffusion at $\approx3$~PeV, thus generating the \knie.
Diffusion and drift of different cosmic-ray nuclei at ultrarelativistic
energies depend on the energy per unit charge $E/Z$. Thus, the position of the
\knie\ for each kind of nucleus is proportional to $Z$. 

Based on the large halo model Kalmykov and Pavlov \cite{kalmykov} have fitted
the primary mass composition in order to obtain best agreement with the
experimental size spectrum of the MSU array.  The resulting energy spectra are
shown in Figure~\ref{kalmykov} for five mass groups as well as the all-particle
spectrum. The flux given by the authors has been normalized to the average
experimental all-particle flux at 1~PeV.  A rigidity dependent cut-off for the
individual element spectra is clearly visible.

Except for the absolute normalization, the shape of all elemental groups is
very similar for this approach and the \modell. Above 30~PeV the iron group is
clearly dominant.  This is also reflected by the pronounced increase of the
mean logarithmic mass, as can be inferred from Figure~\ref{lnamod}. The model
predicts a very strong increase of $\lna$ within a small energy interval.  At
$10^8$~GeV no other model yields such high values of $\lna$.

\subsection{Diffusion in turbulent galactic magnetic fields}

\begin{figure*}\centering
 \epsfig{file=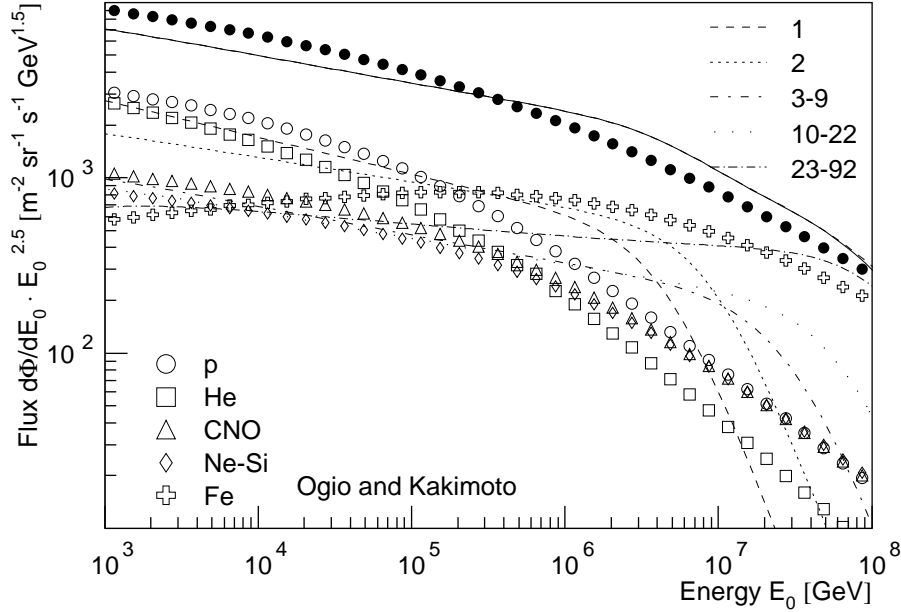,width=12cm}
 \caption{Energy spectra for groups of elements
	  for the model by Ogio and Kakimoto \cite{ogio} (open symbols)
	  and the \modell, see caption of Figure~\ref{stanev}.}
 \label{ogio}
\end{figure*}

Ogio and Kakimoto \cite{ogio} consider the regular magnetic field in the Galaxy
following the direction of the spiral arms. In addition, irregularities of
roughly the same strength are supposed to exist.  It is assumed that the
regular and irregular components have about the same field strength $B\approx
3$~$\mu$G and that both decrease exponentially with a scale hight of 1~kpc. The
scale length of the irregularities is estimated to about $L_{irr}\approx
50$~pc.

The magnetic field lines of the loop and filament structures are open and
perpendicular to the galactic disk. In these structures of the disk the
confinement of cosmic rays is not strong and the particles are leaking rapidly.
The authors assume that the particles leak at this structures with diffusive
motions along the magnetic field lines with a diffusion coefficient
$D_\parallel$. Additionally, outflows with galactic winds with a velocity
$v_g\approx5\cdot10^{-4}$~pc/a are anticipated.  The leakage of cosmic rays
from the Galaxy is described with an one-dimensional advective-diffusion
equation
\begin{equation}
 \frac{\partial n}{\partial t} + v_g \frac{\partial n}{\partial x} =
 \frac{\partial}{\partial x}
 \left( D_{\parallel}\frac{\partial n}{\partial x}\right) +Q \quad.
\end{equation}

The diffusion equation is solved numerically and the residence time $\tau_R$ of
cosmic-ray particles in the galactic disk is obtained.  It is found that
$\tau_R$ depends on the charge of the particles and that it is smaller for
light nuclei. This implies that the average mass of cosmic-ray particles is
expected to increase with energy.  The relative abundances and the spectral
indices for each cosmic-ray component are taken from direct measurements and
the all particle spectrum is normalized to direct measurements at $10^{12}$~eV.

The resulting energy spectra for five groups of elements are presented in
Figure~\ref{ogio} together with the all-particle spectrum obtained.  The
individual components exhibit a relatively smooth \knie\ structure, the bends
extend over several decades in primary energy. The observed \knie\ structure in
the all-particle spectrum is not well reproduced by this model.  The shape of
the average measured energy spectrum (solid line in Figure~\ref{ogio}) is
described only moderately by the model calculations (filled points).

\begin{figure*}\centering
 \epsfig{file=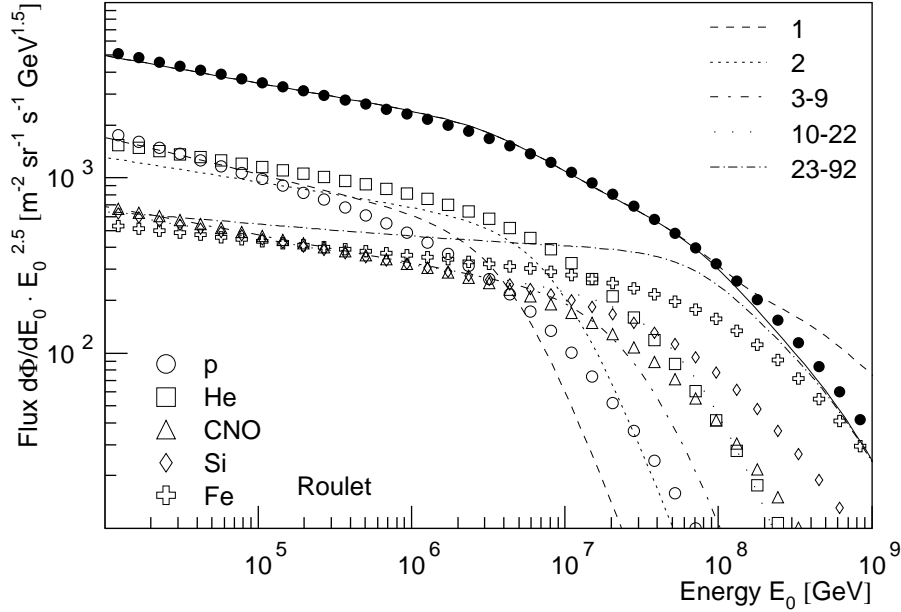,width=12cm}
 \caption{Energy spectra for groups of elements for the model by Roulet 
	  \cite{roulet} and the \modell, see caption of Figure~\ref{stanev}.}
 \label{roulet}	  
\end{figure*}

Since the heavy component clearly dominates above $10^6$~GeV, a relative
strong increase of the mean logarithmic mass with energy is expected, see
Figure~\ref{lnamod}.  At the highest energies ($E_0\approx 10^8$~GeV) values of
$\lna>3$ are obtained.  The increase occurs with an almost constant slope in
the logarithmic plot, while the two other diffusion models considered, exhibit
a distinct change in the \knie\ region.

\subsection{Diffusion and drift}  \label{secroulet}
Similar to the models discussed previously, Roulet et al. \cite{roulet}
consider the drift and diffusion of cosmic-ray particles in the regular and
irregular components of the galactic magnetic field.  Again, a three-component
structure of the magnetic field is assumed.  The regular component is aligned
with the spiral arms, reversing its directions between consecutive arms. This
field (with strength $B_0$) will cause particles with charge $Z$ to describe
helical trajectories with a Larmor radius $R_L=p/(Z e B_0)$.  Secondly, a
random component is assumed. This will lead to a random walk and diffusion
along the magnetic field direction, characterized by a diffusion coefficient
$D_\parallel\propto E^m$. The diffusion orthogonal to the regular magnetic
field is typically much slower, however the energy dependence of $D_\perp$ is
similar to $D_\parallel$.  The third component is the antisymmetric or Hall
diffusion, which is associated with the drift of cosmic rays moving across the
regular magnetic field. The antisymmetric diffusion coefficient is $D_A\approx
r_L c /3 \propto E$.  

In this model, for energies below $Z\cdot E_k$ perpendicular diffusion is the
dominant effect, while for larger energies the drifts are responsible for the
dominant escape mechanism.  The transition between these two regimes is
naturally understood from the different energy dependence of the two diffusion
coefficients $D_\perp$ and $D_A$.  Both coefficients are comparable,
$D_\perp\approx D_A$, at an energy of the order of $Z\cdot E_k$. This generates
a break in the all-particle spectrum.

\begin{figure*}\centering
 \epsfig{file=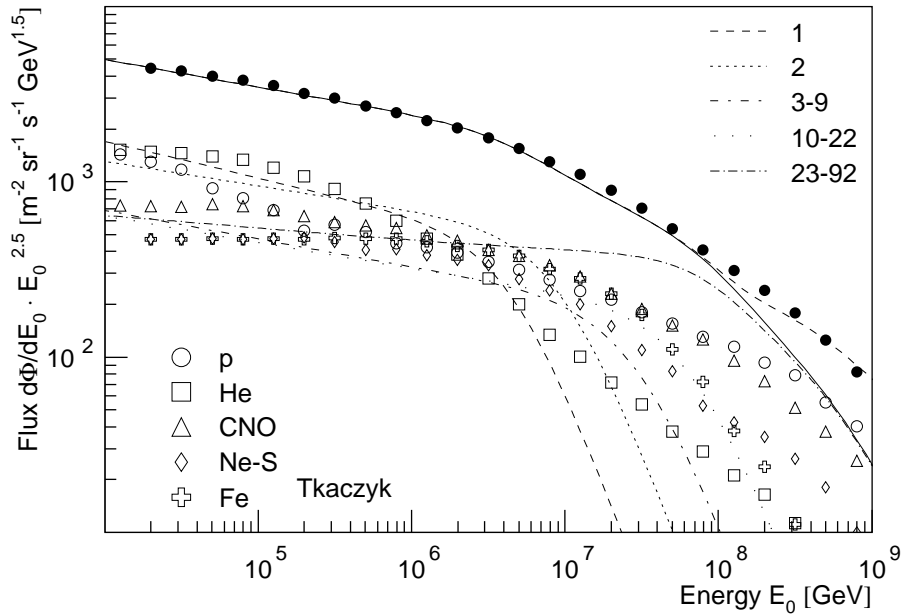,width=12cm}
 \caption{Energy spectra for groups of elements for the model by 
	  Tkaczyk \cite{tkaczyk} (symbols) and the \modell, see caption of
	  Figure~\ref{stanev}.}
 \label{tkaczyk}
\end{figure*}

For the calculations a source spectrum $dQ/dE \propto E^{-\alpha_s}$, with a
constant index for all species $\alpha_s = 2.3$ is assumed.  Below the \knie,
where transverse turbulent diffusion dominates, the observed spectral index
will be $\alpha \approx \alpha_s + 1/3$, while in the drift dominated region
above $E\cdot E_k$, $\alpha\approx\alpha_s+1$ is obtained.  This results in a
rigidity dependent cut-off for the individual elements, as can be seen in
Figure~\ref{roulet}. For the calculations, the cosmic-ray composition has been
normalized to direct measurements at energies below $10^{14}$~eV.  The change
of the slope for individual components $\Delta\alpha\approx2/3$ is softer than
compared to the \modell, where $\Delta\gamma=2.10$ has been obtained.

Nonetheless, the overall agreement between the two approaches is quite good.
The iron component is clearly dominant at the highest energies shown.  This
results in a relatively strong increase of the mean logarithmic mass as
function of energy, as can be seen in Figure~\ref{lnamod}.  At $10^9$~GeV
$\lna\approx 3.75$ is reached.

\subsection{Photo-disintegration and diffusion} 
Interactions of cosmic rays with various background particles are considered in
the next three models described.

Following an idea suggested by Hillas \cite{hillas}, several authors discuss
the possibility of photo-disintegration of nuclei in a dense field of soft
photons around discrete galactic sources, e.g. pulsars. Cosmic rays are assumed
to be magnetically confined near the source and, therefore, accumulate a large
column density on their pass across the photon field. These models are inspired
by the indication in some experiments that the mass composition becomes lighter
beyond the \knie. The \knie\ is explained by the onset of photodisintegration
as well as due to leakage from the Galaxy in the diffusion process in the
galactic magnetic field.

Tkaczyk \cite{tkaczyk} assumes the individual spectra to be described by power
laws with a spectral index of $\gamma_p=-2.75$ for protons and $\gamma_x=-2.55$
for all other nuclei up to iron using the abundances observed at 100~TeV by the
JACEE experiment. The photon background is taken to have a Planck type
distribution with $k_BT=20$~eV. Three processes of energy loss are taken into
account: pair production, pion photo production on nucleons, and
photo-disintegration of nuclei. The magnetic field is assumed to be dominated
by its turbulent component namely 93\% of 31.6~$\mu$T and the radius of the
halo is extended to 15~kpc. Trajectories of particles have been calculated
starting at random positions inside the galactic disk.

\begin{figure*}\centering
 \epsfig{file=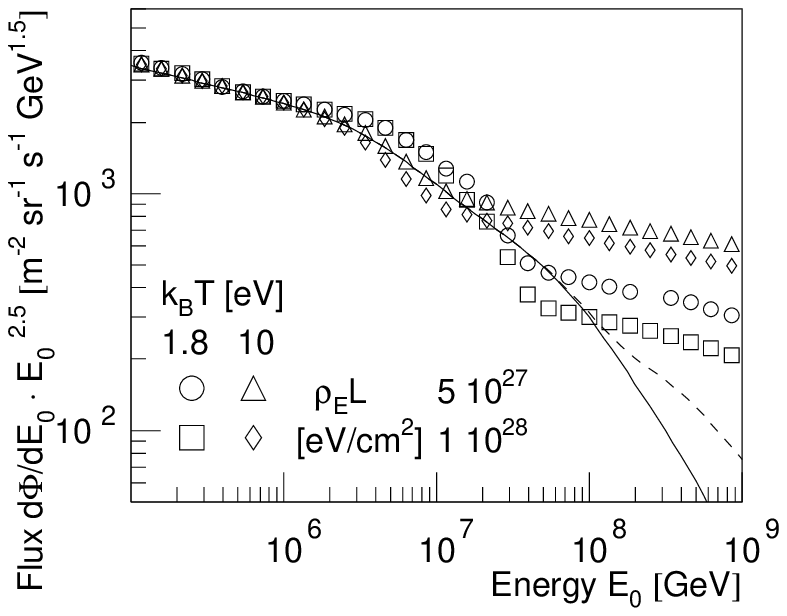,width=\columnwidth}
 \epsfig{file=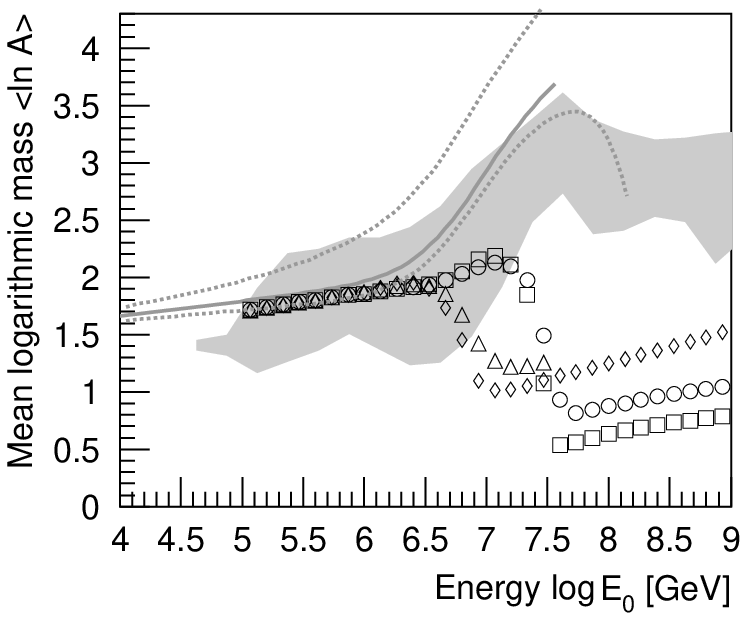,width=0.95\columnwidth}
 \caption{All-particle energy spectrum (left-hand) and mean logarithmic
	  mass versus primary energy (right-hand) for the model by Candia et
	  al. \cite{candia} (symbols), see also captions of
	  Figures~\ref{lnamod} and \ref{stanev}. Results for different values
	  for the temperature $T$ and the column density $\rho_EL$ of the
	  photon field are shown.}
 \label{candia}	  
\end{figure*}

The energy spectra obtained for elemental groups are presented in
Figure~\ref{tkaczyk}. The energies as given in \cite{tkaczyk} have been
rescaled by $\delta_E=-12\%$ in order to match the all-particle spectrum
(filled circles) with the average measured values (solid line). The shape of
the all-particle spectrum obtained is in good agreement with the measured
values. The cut-off behavior for individual element groups is softer than in
the \modell.  The steepening in the spectrum of primary protons at $\approx
10^5$~GeV is due to their leakage from the Galaxy. In the region between 1~PeV
and 30~PeV the \knie\ in the all-particle spectrum is explained as due to
photo-disintegration of nuclei and due to leakage.  The photo-disintegration
process delivers a large contribution of secondary protons which leads protons
to become the dominant mass group above $10^8$~GeV.  

The predicted mean logarithmic mass, shown in Figure~\ref{lnamod}, exhibits a
slight increase between $10^4$ and $10^7$~GeV in fair agreement with the
measurements. Due to the strong contribution of secondary protons, $\lna$
decreases above $10^7$~GeV in contrast to the further increase indicated by the
measured values.

\subsection{Photo-disintegration by optical and UV photons} 
Photodisintegration of cosmic-ray nuclei in dense photon fields around compact
sources is also discussed in the model by Candia, Epele, and Roulet
\cite{candia}. They investigate in detail the potential influence of cosmic
rays passing through columns of optical and soft UV photons in the source
region. They vary the column density and $k_BT$ in the Planck spectrum to study
the effect on the \knie\ structure. Increasing the temperature $T$ shifts the
\knie\ towards lower energies whilst increasing the column density $\rho_E L$
intensifies the steepening at the \knie. The authors conclude that a
comparatively small column density of about $5\cdot10^{27}$~eV/cm$^2$ and a
soft photon spectrum with $k_BT\approx2$~eV describes the \knie\ structure best
and can, at least partly, be responsible for it.

The total energy spectrum as calculated by the authors and the appertaining
mean logarithmic mass are plotted in Figure~\ref{candia}. The spectra at the
source are assumed to be featureless extrapolations of the spectra below the
\knie. Spectra for two values of the Planck temperature $k_BT=1.8$~eV and
$k_BT=10$~eV as well as two column densities $\rho_EL=5\cdot10^{27}$~eV/cm$^2$
and $\rho_EL=10^{28}$~eV/cm$^2$ are presented. 
The main mechanism of energy loss for nuclei with energy $E_0<10^{18}$~eV is
the process of photo-disintegration.  If the typical energy of the photons is
in the optical range (1-10~eV), the photo-disintegration of cosmic-ray nuclei
with mass $A$ will start to be efficient at energies $E_0\ge A\cdot10^{15}$~eV.
The appearance of the \knie\ is ultimately a threshold effect.

Up to $10^7$~GeV the shape of the all-particle spectrum is compatible with the
average measured flux. Candia et al. remark that excessive fluxes at energies
above $10^8$~GeV are not troublesome since leakage from the Galaxy or different
efficiencies in the acceleration mechanisms have not been considered in the
present model, but they can not be disregarded.

The mean logarithmic mass as derived from the present model is shown as
function of energy in Figure~\ref{candia} on the right-hand panel. Again,
results taking into account different temperatures and column densities are
presented.  The photo-disintegration of heavy nuclei yields a strong decrease
of $\lna$ above the \knie. The extreme light composition above $10^7$~GeV
is significantly below the observed values.

\subsection{Neutrino interactions in the galactic halo}  

Dova, Epele and Swain \cite{dova} link the origin of the \knie\ to a
"GZK-cut-off like effect" of cosmic rays interacting with massive neutrinos in
the galactic halo. The average number density $n_\nu=337$/cm$^3$ of
standard-model neutrinos with mass $m_\nu<1$~MeV as predicted by Big Bang
cosmology is assumed to be strongly increased due to gravitational clustering
in galaxies to a value $n_\nu=1.4 \cdot 10^8$/cm$^3$.  In addition, a magnetic
dipole moment $\mu_\nu = 5.4\pm0.6 \cdot 10^{-6}~\mu_B$\footnote{$\mu_B$: Bohr
magneton} is adopted for massive neutrinos in order to increase the
cross-section for the inelastic scattering of nucleons on the neutrino
background.  

Pion production with a threshold energy of $E_p=3$~PeV is anticipated via the
processes $p+\nu\rightarrow\nu+\Delta$ and $\Delta\rightarrow p+\pi$.  This
fixes the neutrino mass at $m_\nu=100$~eV. The relative heavy mass is
attributed to muon or tau neutrinos, but exotic objects like hypothetical
dark matter particles are named as candidates as well. These particles are
proposed to be inside a spheroidal galactic halo with a radius of 10~kpc.  The
cosmic-ray spectrum is modeled as a two component mixture ($\approx 60\%$
protons + 40\% iron) with a spectral index $\gamma=-2.8$ for both components.
The propagation is described by a diffusion model, taking into account the
galactic magnetic field and a residence time of about $3\cdot 10^8$~a,
independent of energy.

\begin{figure}
 \epsfig{file=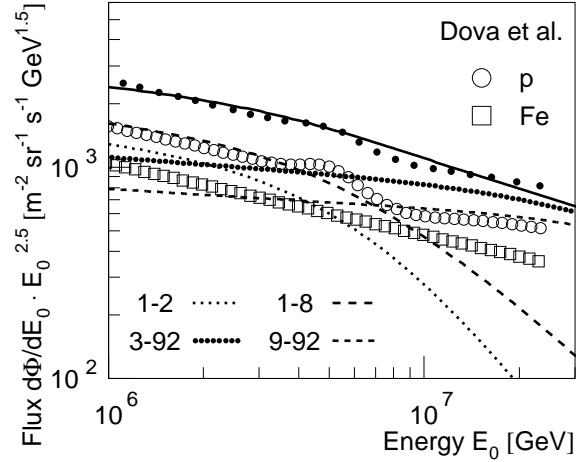,width=\columnwidth}
 \caption{Energy spectra for groups of elements for the model by
	  Dova et al. \cite{dova} (symbols) and the \modell\ (lines).
	  The numbers indicate the range of nuclear charge used for the
	  elemental groups, the all-particle spectrum is represented by the
	  filled points and the solid line, respectively.}
 \label{dova}	  
\end{figure}

The spectra calculated for the two groups "proton" and "iron" are given in a
small energy range from 1~PeV to 30~PeV in the publication and are shown in
Figure~\ref{dova}.  The energy scale of the model has been renormalized by
$\delta_E=-6\%$ in order to fit the average all-particle flux of the
measurements (solid line).  In the figure the two components are compared with
the \modell\ in two ways: the light component is taken as the sum of protons
and helium on the one hand and as the sum for $1\le Z\le8$ on the other hand.
The remaining elements up to $Z=92$ are used for the heavy component in each
case.  Neither set of spectra is compatible with the calculations of Dova et
al. as can be inferred from the figure. The flux of light elements above
the \knie\ is overestimated by the model.

The mean logarithmic mass, as calculated from the two groups is
almost constant, with an average value of about 2 and a slight modulation
around $10^{6.75}$~GeV, see Figure~\ref{lnamod}.  At the highest energies
shown, the trend towards a light mass composition is not supported by the
experimental values.

\subsection{New physics in the atmosphere}
A completely different reason for the \knie\ is discussed by Kazanas and
Nicolaidis \cite{kazanasall}. They investigate nucleon interactions hitherto
unaccounted for in  air showers, which transfer energy into particles not
observable (or not yet observed) by air shower experiments. Candidates for new
physics are supersymmetry, with the lightest supersymmetric particle carrying
away the missing energy, technicolor models which produce techni-hadrons not
observed by present experiments, or models in dimensions $>4$ lowering the
characteristic energy for gravitation to TeV energies and thus producing
graviton energy carried away by "invisible" particles. Hence, the true energy
of the shower inducing particle is underestimated in air shower measurements
above the threshold energy. Being a threshold effect the new interaction
entails a relatively sharp \knie\ structure.

The conventional proton-proton total cross-section rises slowly with energy.
In the model it is considered to have a constant value of $\sigma_0=80$~mbarn.
The additional cross-section $\sigma_n(E)$ describes the proposed new type of
interaction, rising strongly with energy.  At energies above the threshold a
fraction of protons interacts with the probability
$P_n(E)=\sigma_n(E)/[\sigma_0(E)+\sigma_n(E)]$ through the new channel. Hence,
more and more energy is missing in the measured components.

\begin{figure}[bt]
 \epsfig{file=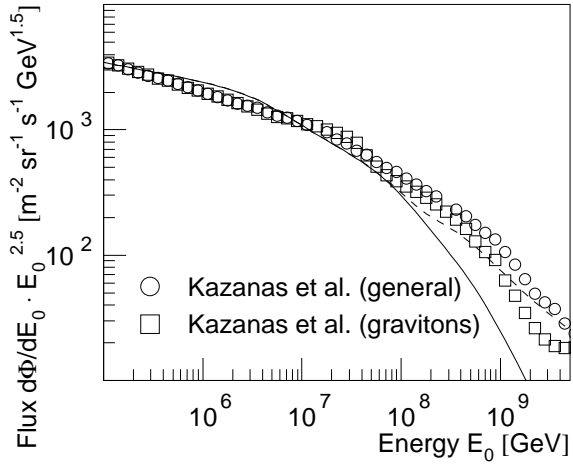,width=\columnwidth}
 \caption{All-particle energy spectrum according to calculations by 
	  Kazanas \& Nicolaidis for a general model \cite{kazanasall} and for
	  graviton production \cite{kazanasgrav} (symbols). The average
	  measured all-particle flux is shown by the solid line and above
	  $10^8$~GeV as dashed line.}
 \label{kazanas}	  
\end{figure}

As primary all-particle energy spectrum the authors assume a single power law
with a cut-off at an energy $E_0$ of the form $I=E^{-\gamma} \exp(-E/E_0)$ with
$\gamma=2.75$ and $E_0=10^{18.5}$~eV. In addition, at energies above
$10^{18}$~eV, an extragalactic component is introduced. The all-particle
spectrum above the atmosphere is modulated through the proposed new
interactions.  The resulting apparent spectrum, reconstructed from observations
below the atmosphere is plotted in Figure~\ref{kazanas} (circles). The model
produces a spectrum with a shape similar to the average measured values, the
\knie\ occurs at a relative high energy above $10^7$~GeV. The authors point out
that the present model is very simple. Hence, some deviations between
measurements and the predicted spectrum are possible.

In a second article Kazanas and Nicolaidis perform more detailed calculations
of the cross-section in the case of graviton radiation \cite{kazanasgrav}.  As
new physics in the interactions low scale gravitation is discussed. The
graviton is considered to propagate not only in the usual four dimensions but
in an additional compactifyed large dimension $\delta$.  Thus the fundamental
(Planck) scale of gravity $M_f$ is reduced to TeV energies.  Any scattering
process at center-of-mass energies $E>M_f$ should be accompanied by abundant
graviton production, i.e.  collinear bremsstrahlung of gravitons.  The
cross-section for graviton production in $pp$ collisions is calculated to be
$\sigma_n\propto (\sqrt{s}/M_f)^{2+\delta}$. A fit to the measured cosmic-ray
spectrum yields $M_f\approx8$~TeV and $\delta=4$. This indicates a strong
increase of the graviton production cross section as function of energy. Even
at energies just above $M_f$, graviton radiation will be an important process.

Again, as primary spectrum a single power law has been assumed and the apparent
spectrum has been calculated. It is shown in Figure~\ref{kazanas} (squares).
The spectrum is very similar to the one obtained from the more general
consideration for different processes as mentioned above.

\section{Discussion}

\begin{table*}
 \caption{Synopsis of all models discussed.}
 \label{sumtab}
 \renewcommand{\arraystretch}{1.0}
 \begin{tabular}{ll}\hline
  Model&Author(s)\\ \hline
  Source/Acceleration:&\\
  ~~~Acceleration in SNR       & Berezhko \& Ksenofontov \cite{berezhko}   \\
  ~~~Acceleration in SNR + radio galaxies & Stanev et al. \cite{stanev}  \\
  ~~~Acceleration by oblique shocks  & Kobayakawa et al. \cite{kobayakawa}\\
  ~~~Acceleration in variety of SNR& Sveshnikova \cite{sveshnikova}\\
  ~~~Single source model & Erlykin \& Wolfendale \cite{wolfendale}  \\
  ~~~Reacceleration in the galactic wind& V\"olk and Zirakashvili \cite{voelk}\\
  ~~~Cannonball model & Plaga                 \cite{plaga}  \\
  Propagation/Leakage from Galaxy:&\\
  ~~~Minimum pathlength model & Swordy  \cite{swordy}  \\
  ~~~Anomalous diffusion model & Lagutin et al. \cite{lagutin}  \\
  ~~~Hall diffusion model & Ptuskin et al. \cite{ptuskin}, Kalmykov \& Pavlov
                           \cite{kalmykov}  \\
  ~~~Diffusion in turbulent magnetic fields& Ogio and Kakimoto \cite{ogio}  \\
  ~~~Diffusion and drift                   & Roulet et al \cite{roulet} \\
  Interactions with background particles:&\\    
  ~~~Diffusion model + photo-disintegration & Tkaczyk \cite{tkaczyk} \\
  ~~~Interaction with neutrinos in galactic halo & Dova et al. \cite{dova}\\
  ~~~Photo-disintegration (optical and UV photons)& Candia et al. \cite{candia}
     \\ 
  New interactions in the atmosphere:&\\   
  ~~~Gravitons, SUSY, technicolor& 
          Kazanas \& Nicolaidis \cite{kazanasall,kazanasgrav}\\
     \hline
 \end{tabular}
\end{table*}

All models discussed in this article are summarized in Table~\ref{sumtab}.
Most of them yield similar all-particle spectra. On the other hand, the
individual element spectra are quite different in some models. Some predict a
very weak \knie\ structure, like e.g. the acceleration in cannon balls by Plaga
and the anomalous diffusion described by Lagutin et al, while others obtain a
steep falling flux at the \knie, an example is the diffusion model by Kalmykov
and Pavlov. The results from indirect measurements seem to favor a relative
steep cut-off, see section~\ref{pgsect}.

Most models predict featureless spectra with a simple shape, however, some
scenarios yield distinct features, which may have different theoretical
origins.  Kalmykov and Pavlov as well as Swordy consider a diffusion model and
a Leaky Box model, respectively and obtain bumps at the individual \knie\
positions. Consequently, they find a dip at \knie\ energies in the mean
logarithmic mass.  The approaches by Sveshnikova (acceleration in supernova
remnants) as well as V\"olk and Zirakashvili (reacceleration in the galactic
halo) lead both to a twofold shape of the element spectra.  Yet, also common
features can be recognized, like the relative sharp kinks in the spectra as
found by Erlykin and Wolfendale as well as by Stanev et al., both models
describe acceleration in supernova remnants. The distinct shape leads in both
cases to structures in the $\lna$ distribution. 

The review has shown further that even similar conceptions lead to $\lna$
values covering the whole range depicted in Figure~\ref{lnamod}.  As an
example, acceleration in supernova remnants yields $\lna$ values at the lower
edge of the experimental values (Berezhko and Ksenofontov) and also at the
upper border (Kobayakawa et al.).

Unfortunately, the experiments give no conclusive energy spectra
for elemental groups (Figures~\ref{kascade} - \ref{hegra}) and the mean
logarithmic mass covers a relatively wide range, see Figure~\ref{lnamod}.  Most
astrophysical informations available from air shower measurements depend on the
model used to describe the interactions in the atmosphere.  Hence, an
interpretation and evaluation of the data should be treated with care and it is
very challenging to characterize in a quantitative way the best model to
describe the origin of the \knie.

In models which introduce new type of interactions in the atmosphere, like the
one just discussed by Kazanas and Nicolaidis or the approaches by Nikolsky et
al. \cite{nikolsky} or by Petrukhin \cite{petrukhin}, the threshold for a new
type of interaction depends on the energy per nucleon. Thus, such models
ultimately arrive at a mass dependent cut-off for individual element spectra.
The investigations of the KASCADE group however seem to indicate a rigidity
dependent cut-off for individual elements
\cite{ulrich,ulrichisvh,rothisvh,roth,taup}.  Also with the phenomenological
approach of the \modell\ a rigidity dependent approach is favored
\cite{polygonato}.  

In addition, the simultaneous observations of a \knie\ in the electromagnetic,
muonic and hadronic component by the KASCADE group \cite{ulrich,emhknie,hknie}
yield no major inconsistencies between the different air shower components.
Therefore, a proposed new interaction would have to "hide" the energy in such a
way, that no inconsistencies occur, when the measurements of different shower
components are interpreted with standard particle physics. From the present
experimental results new particle physics for high-energy interactions in the
atmosphere seems not to be a likely explanation for the \knie. 

The interaction of cosmic rays with background particles like neutrinos or
photons during their propagation causes fragmentation of heavy nuclei and the
production of a large amount of secondary light nuclei, preferable protons.
Models based on such an approach, like the models by Tkaczyk, Dova et
al.\footnote{Dova et al give their results only for a very limited range in
primary energy, but above $10^7$~GeV a trend towards an extremely light mass
composition can be recognized.}, and Candia et al. yield a very light mean
logarithmic mass above \knie\ energies.  On the other hand, the measurements
indicate an increase of $\lna$ with energy, see Figure~\ref{lnamod}, thus
disfavoring interactions with background particles in the Galaxy as an
explanation for the \knie. In addition, the cross-sections for such
interactions depend on the energy per nucleon. This yields a mass dependent
cut-off for individual element spectra, which seems not to be favored by the
measurements as mentioned above.

Numerous theories consider diffusive propagation of cosmic rays through the
interstellar medium and leakage from the Galaxy as an explanation of the \knie.
The escape probability from the Galaxy in these models depends
on the nuclear charge $Z$ and a rigidity dependent cut-off behavior is
obtained for the individual element spectra.
These models yield a more or less strong increase of the mean logarithmic mass
above the \knie\ as consequence of the leakage starting with light nuclei.  The
change of the spectral slope at the \knie\ is not very distinct in the Leaky
Box model by Swordy as well as in the anomalous diffusion model by Lagutin et
al. Consequently, only a modest increase of the mean logarithmic mass is
obtained for energies above the \knie, at the highest energies the predicted
$\lna$ values are at the lower edge of the measurements.

Propagation of cosmic rays in the Galaxy, taking into account contributions from
the regular magnetic field, a random component and antisymmetric diffusion are
described by Kalmykov and Pavlov, Ogio and Kakimoto, as well as Roulet et al.
Although they consider similar scenarios, quite different energy spectra are
obtained.  Ogio and Kakimoto derive a very smooth transition in the \knie\
region, while Kalmykov and Pavlov as well as Roulet et al. calculate a cut-off
behavior similar to the measurements.  The mean logarithmic mass obtained for
the three models covers a wide range,  and around $10^6$~GeV a significantly
different behavior for the individual models is found.  On the other hand,
they all result in similar values of $\lna\approx3$ around $10^{7.5}$~GeV, well
compatible with the measurements.

Several models attribute the \knie\ to the acceleration process of cosmic rays.
In the cannonball model described by Plaga, the spectral slopes change only by
about $\Delta\gamma\approx0.3$ at the \knie\ and protons are the dominant
nuclei even above the \knie. Hence, the resulting mean logarithmic mass is
almost a constant function of energy. It is the weakest energy dependence of
all models discussed.  In the single source model by Erlykin and Wolfendale
heavy nuclei dominate above the \knie\ and the $\lna$ values are in fair
agreement with the measurements.  This can also be stated for the concept of
V\"olk and Zirakashvili \cite{voelk}, which allows for (re)acceleration of
cosmic rays up to energies of the ankle.

Diffusive shock acceleration in supernova remnants is considered by Berezhko
and Ksenofontov, Stanev et al., Kobayakawa et al., and Sveshnikova.  Different
variations of the basic theory of Fermi acceleration have been applied.  Due to
the finite lifetime of the shock front the maximum energy reached is limited.
Typical values of $Z\cdot (0.1 - 3)$~PeV are attained.  Kobayakawa et al. could
increase the maximum energy reached in their calculations by the investigation
of special magnetic field configurations.  Sveshnikova takes into account
recent supernova observations, which indicate the existence of very energetic
explosions (hypernovae). In their high-velocity shock fronts particles can
effectively be accelerated to high energies.  The spectra obtained by the
latter two scenarios are similar to the measurements and the \modell.

The \modell\ is purely phenomenological. The spectra obtained are most
compatible with the acceleration in supernova remnants as described by
Sveshnikova, especially the heavy enriched case, and diffusive propagation
through the Galaxy as discussed by Kalmykov and Pavlov as well as Roulet et al.
Hence, these models could be a good theoretical foundation of the
phenomenological ansatz.

Most models discussed represent principle ideas, dealing with individual
aspects of cosmic-ray acceleration, propagation, or interactions.  Most
probably, combinations of this basic processes are needed to describe nature
realistically and more precise data will be necessary for a fine tuning of the
free parameters in the models.  These parameters have been adjusted to
different sets of experimental results in the individual models.  Most likely
some of their predictions change, if a different set of experimental results is
used.

At present, the experimental results do not allow to select a single approach
as the best model to describe nature. The origin of the \knie\ in the
all-particle energy spectrum is not yet explained.  However, within the
boundaries given by the experiments, acceleration in supernova remnants and
diffusive propagation through the Galaxy seem to be very attractive models to
understand the origin of the \knie.

\section*{Acknowledgment} 
The author would like to thank A.A.~Lagutin and E.~Roulet for providing
numerical data for their models.  
Valuable discussions with J.~Engler, N.N.~Kalmykov, and K.-H.~Kampert are
acknowledged.

\end{document}